\pdfoutput=1
\documentclass[11pt]{article}

\usepackage[affil-it]{authblk} 
\usepackage[authoryear]{natbib}
\usepackage{gensymb}
\usepackage{amsmath}
\usepackage{amsfonts}
\usepackage[T1]{fontenc}
\usepackage{courier}
\usepackage{amssymb}
\usepackage{graphicx}
\usepackage{float}
\usepackage{booktabs}
\usepackage[a4paper, top=1in,bottom=1in,left=1in,right=1in,marginparwidth=0cm]{geometry}
\usepackage[title]{appendix}
\usepackage{pdflscape}
\usepackage{longtable}
\usepackage{setspace}
\usepackage{adjustbox}
\usepackage{array}
\usepackage{caption}
\usepackage{chngcntr}
\usepackage{mathtools}
\usepackage{lineno}
\usepackage{makecell}
\usepackage{hyperref}
\hypersetup{colorlinks=true,allcolors=blue}
\usepackage{longtable}

\newcolumntype{R}[2]{%
    >{\adjustbox{angle=#1,lap=\width-(#2)}\bgroup}%
    l%
    <{\egroup}%
}
\newcommand*\rot{\multicolumn{1}{R{45}{1em}}}

\renewcommand{\eqref}[1]{Eq.\,\ref{#1}}

\title{Biological barriers to forest pest invasions: A novel host tree slows mountain pine beetle range expansion}

\author[1,*]{Evan C. Johnson}
\author[2]{Antonia Musso}
\author[3]{Catherine Cullingham}
\author[4,5]{Mark A. Lewis}
\affil[1]{Mathematical and Statistical Sciences; University of Alberta; Edmonton, Alberta, Canada}
\affil[2]{Biological Sciences; University of Alberta; Edmonton, Ontario, Canada}
\affil[3]{Department of Biology; Carleton University; Ottawa, Ontario, Canada}
\affil[4]{Department of Mathematics and Statistics; University of Victoria; Victoria, British Columbia, Canada}
\affil[5]{Department of Biology; University of Victoria; Victoria, British Columbia, Canada}
\affil[*]{Corresponding author: Evan Johnson, ecjohns1@ualberta.ca}

\date{} 

\begin{document}

\maketitle 

\newpage

\tableofcontents 

\newpage 


\section*{Abstract} 

Following widespread outbreaks across western North America, mountain pine beetle recently expanded its range from British Columbia into Alberta. However, mountain pine beetle's eastward expansion across Canada has stalled unexpectedly, defying predictions of rapid spread through jack pine, a novel host tree.  This study investigates the underlying causes of this deceleration using an integrative approach combining statistical modeling, simulations, and experimental data. We find that the slow spread is primarily due to mountain pine beetle's difficulty in finding and successfully attacking jack pine trees, rather than issues with reproduction or larval development. The underlying mechanism impeding beetle range expansion has been hypothesized to be lower pine volumes in eastern forests, which are primarily a consequence of lower stem density. However, our analysis suggests that jack pine's phenotype itself is the primary impediment. We propose that jack pine's smaller size, thinner phloem, and lower monoterpene concentrations result in weaker chemical cues during the host-finding and mass-attack stages of MPB's life cycle, ultimately leading to fewer successful attacks. These findings suggest a reduced risk of further eastward spread, but should be interpreted cautiously due to enormous policy implications and the inherent limitations of ecological forecasting.

\newpage

\section{Introduction} \label{Introduction}

The most recent outbreak of the mountain pine beetle (MPB; \textit{Dendroctonus ponderosae} Hopkins), occurring roughly from 2000 to 2015, was the largest bark beetle outbreak ever recorded \citep{taylor2006forest}. Spanning across western North America from Arizona to Yukon, this hyperepidemic killed up to 8 million hectares of pine trees \citep{meddens2012spatiotemporal}, and killed $>50\%$ of merchantable pine in British Columbia alone \citep{BCFLNRO2016}. The outbreak caused economic harm \citep{corbett2016economic, abbott2009mountain}, contributed to carbon emissions via wood decomposition \citep{kurz2008mountain}. The outbreak had numerous impacts on hydrological functioning \citep{schnorbus2011synthesis, redding2008mountain} and forest ecosystems (\citealp{dhar2016aftermath}, and sources therein). Fire suppression and climate change and are believed to be major drivers --- MPB thrived due to the weakened defenses of drought-stressed trees and reduced beetle mortality from milder winters \citep{carroll2006impacts, alfaro2009historical, creeden2014climate}. Fire suppression led to a buildup of large-diameter pines, which are favored by MPB \citep{taylor2006forest}.


The recent outbreak was so severe that MPB expanded its historic range in a northeasterly direction, reaching northern British Columbia, the Yukon, the Northwest Territories, and most notably, Alberta \citep{nealis2014risk}. In central British Columbia circa 2005, large numbers of MPB flew above forest canopies --- a behavior hypothesized to occur when beetles are abundant and suitable trees are scarce \citep{carroll2003bionomics, bleiker2019risk}.  These beetles were carried by windstorms across the Rocky Mountains, overcoming a geographical barrier that had historically confined their range \citep{safranyik1989empirical, safranyik1992dispersal, jackson2008radar, de2012breach}.  There is a nearly contiguous path of pine trees across Canada’s boreal zone, so MPB’s arrival in Alberta threatened forests spanning to the east coast. 

\begin{figure}[H]
\centering
\includegraphics[scale = 0.8]{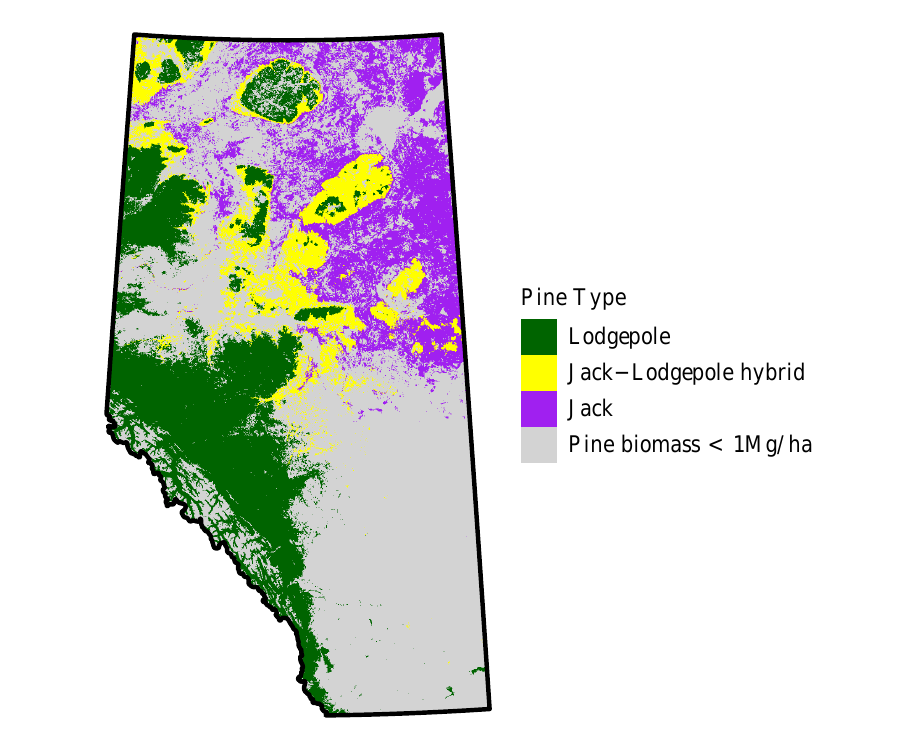}
\caption{Lodgepole pine inhabits western Alberta, while jack pine inhabits eastern Alberta. There is a large zone in central Alberta where the two species hybridize. The pine species data comes from \citet{cullingham2012characterizing}. The pine biomass data comes from \citet{beaudoin2014mapping}.}
\label{fig:pine_in_alberta}
\end{figure}

MPB's ability to spread across Canada is contingent on its ability to spread through a novel host tree, jack pine (\textit{Pinus banksiana}). Jack pine inhabits eastern Alberta, whereas lodgepole pine (\textit{Pinus contorta}) --- MPB's primary host throughout the Rocky Mountains and the Pacific Northwest --- inhabits western Alberta (Fig. \ref{fig:pine_in_alberta}). Initially, the dominant view was that MPB would rapidly spread through jack pine: a working group of MPB researchers concluded that ``There are therefore no known biological impediments to the spread and establishment of the mountain pine beetle through the boreal zone'', and that ``The dynamic behavior of beetles on an invasion front, including the rate of spread, will not differ from observed behavior in similar situations within the historic range'' \citep{nealis2008risk}.

\begin{figure}[H]
\centering
\makebox[\textwidth]{\includegraphics[scale = 0.75]{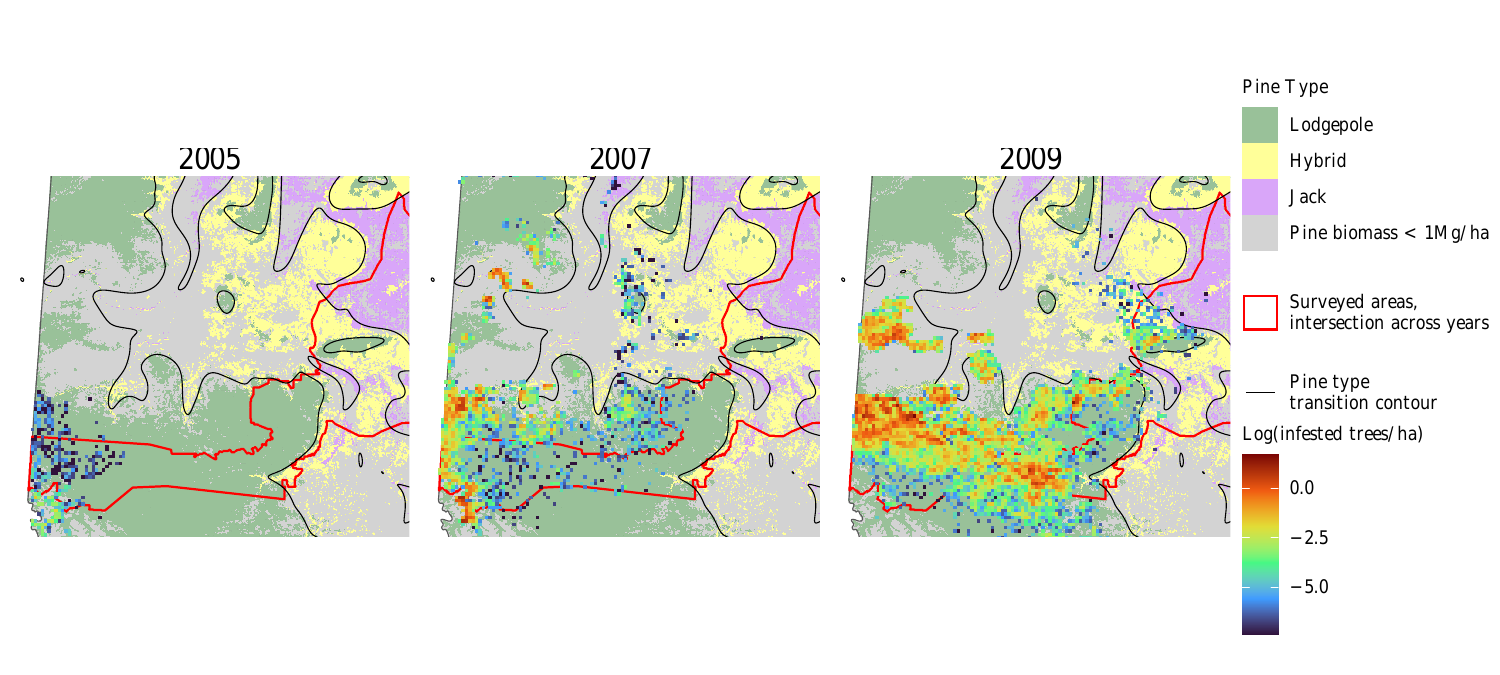}}
\caption{Mountain pine beetle rapidly spread across western Alberta from 2005 to 2009. Infestation data comes from Heli-GPS surveys.}
\label{fig:busy_raster_early}
\end{figure}

\begin{figure}[H]
\centering
\includegraphics[scale = 1]{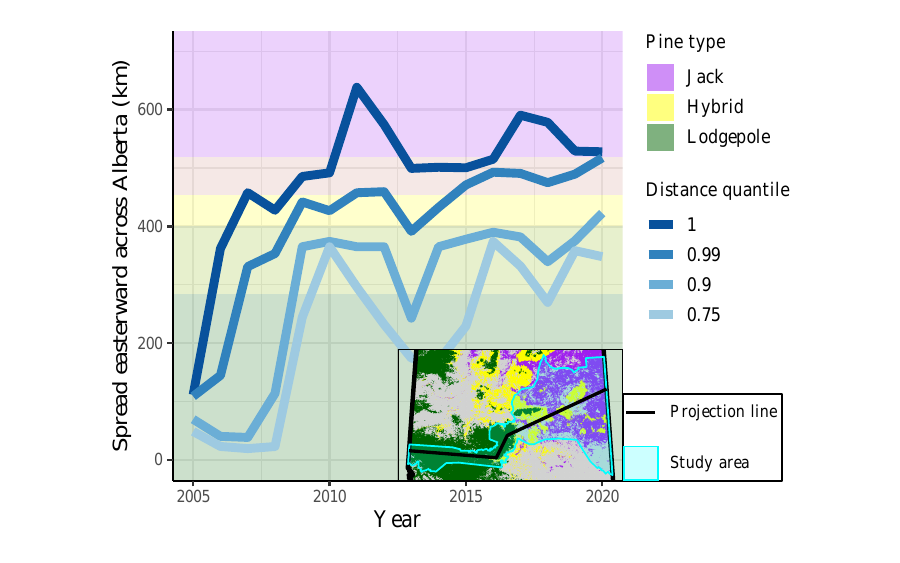}
\caption{Mountain pine beetle's range expansion slowed down significantly around 2009--2010. Spread distance is summarized by projecting infestations onto a single dimension: a line running through the center of the consistently surveyed area (methods in Section \ref{simulations}).}
\label{fig:data_spread_w_inset}
\end{figure}

Mountain pine beetle's range expansion took several unexpected turns (Fig. \ref{fig:busy_raster_early}). Between 2004 and 2006, MPB spread more rapidly than anticipated, quickly reaching areas where lodgepole pine and jack pine hybridize. Most notably, MPB leaped over 220 km eastward in 2006 \citep{cooke2017predicting}. From 2007 to 2009, the spread of MPB continued slightly farther eastward, and large MPB populations developed in lodgepole pine forests around Marten Mountain and directly south of Lesser Slave Lake, areas adjacent to hybrid and jack pine forests. The first successful mountain pine beetle infestation in jack pine, where offspring were produced, was observed in 2009 \citep{cullingham2011mountain}. However, MPB's range expansion then stagnated. For more than a decade, MPB has been unable to proliferate through the jack pine forests of eastern Alberta (Fig. \ref{fig:data_spread_w_inset}).

Understanding MPB's life history is crucial to understanding MPB's decelerating range expansion. Adult beetles synchronously emerge from their natal trees in July or August \citealp{powell2009connecting, bleiker2016flight}. During an initial \textit{host finding/acceptance} stage, females rely on a variety of cues (olfactory, gustatory, and tactile) to locate a suitable host tree \citep{reid1963biology, raffa1982gustatory, raffa2001mixed}; they select a tree by boring into it, but the tree fights back by exuding toxic resin. The pioneering females emit \textit{trans}-verbenol, an aggregation pheromone which attracts additional beetles \citep{pitman1969aggregation, pureswaran2000dynamics}. Arriving males emit \textit{exo}-brevocomin, an aggregation pheromone that attracts additional females \citep{rudinsky1974antiaggregative}. These aggregation pheromones, along with attractive host volatiles, lead to a \textit{mass attack} where swarms of beetles concentrate on a single tree \citep{raffa1983role}. Once the tree's resin reserves are exhausted, females begin colonization by excavating galleries and laying eggs. Larvae then eat the tree's phloem throughout autumn and overwinter with the help of cryoprotectants (\citealp{regniere2007modeling}). They metamorphose in the spring or summer, thus completing their univoltine life cycle.

\begin{figure}[H]
\centering
\makebox[\textwidth]{\includegraphics[scale = 0.85]{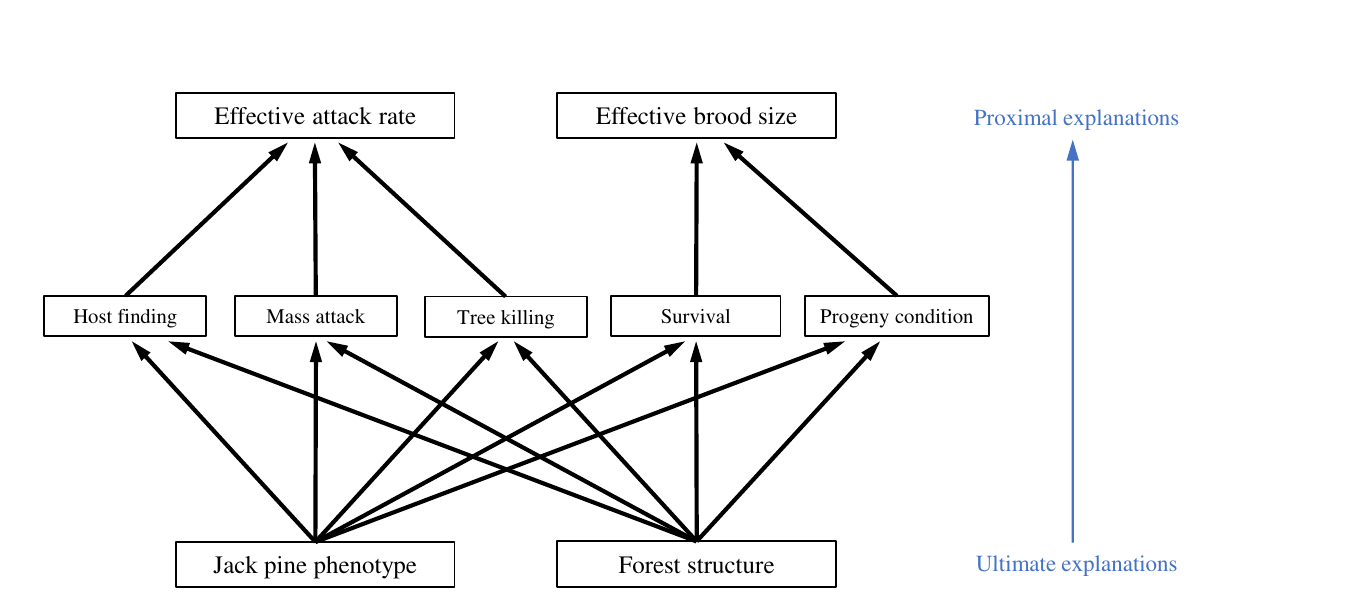}}
\caption{A hierarchy of explanations for MPB's slowed range expansion. The top level consists of model-based constructs: the \textit{effective attack rate} and the \textit{effective brood size}. These encapsulate the life cycle stages in the middle layer. The bottom layer contains ultimate explanations: species-level differences between lodgepole pine and jack pine, and environmental/structural differences between the forests of western and eastern Alberta.}
\label{fig:explanation_hierarchy1}
\end{figure}

A hierarchical framework can be used to organize the numerous explanations for MPB's limited incursion into jack pine forest (Fig. \ref{fig:explanation_hierarchy1}). At the highest level, there are two proximal explanations: (1) a poor effective attack rate in jack pine forests, and (2) a reduced effective brood size. Here, the concept of an \textit{effective attack rate} encapsulates MPB's ability to find and select hosts, mount mass attacks, and overcome tree defenses. The concept of an effective brood size encapsulates per capita fecundity (oviposition), intraspecific competition among larvae, overwintering mortality, and sub-lethal effects of climate and tree defenses which may affect MPB fitness. 

At the bottom level of the hierarchy, there are two ultimate explanations for MPB's slow spread: (1) some species-specific properties of jack pine, and (2) the typical forest structure associated with jack pine in eastern Alberta, i.e., reduced canopy closure, smaller trees, and lower connectivity \citep{burns1990silvics, bleiker2019risk}. The detailed biological mechanisms are conceptualized as arrows between the bottom and middle levels of the hierarchy in Figure \ref{fig:explanation_hierarchy1}. For example, MPB may struggle with host-finding in jack pine forests due to lower levels of $\beta$-phellandrene in jack pine \citep{clark2014comparison}, the only known host monoterpene which attracts beetles in the absence of aggregation pheromones \citep{miller1990beta}. This species-level property of jack pine --- its unique semiochemical profile --- could negatively impact MPB's host-finding abilities, thus negatively affecting the effective attack rate.


We focus on two key contrasts --- effective attack rate versus effective brood size, and pine species identity versus forest structure --- because they are amenable to analysis with models. Solving the puzzle of MPB's slowed range expansion requires an integrative approach that considers all types of evidence. Thus, we consider 1) bolt experiments, where MPB individuals are introduced into cross-cut segments of tree trunks; 2) statistical models, which aim to associate MPB demography with various factors, including pine species identity; and 3) circumstantial evidence, i.e., clues from small-scale studies that focus on a single aspect of MPB biology --- typically behavior or pheromone-based communication. 


Bolt experiments are crucial for determining the effective brood sizes of MPB in lodgepole pine and jack pine. Adult beetles are introduced to bolts (i.e., a short log section) either by inserting harvested beetles into pre-drilled holes in the bark (e.g., \citealp{rosenberger2017cold}), or by catalyzing a mass-attack by placing bolts next to mass-attacked trees (e.g., \citealp{burke2016influence}), or releasing harvested beetles next to the lower bole (e.g., \citealp{musso2023pine}). Typically, the bolts are held in cold storage to mimic the overwintering period, and later warmed to room temperature, allowing beetle development to continue. Beetle performance is measured using several metrics, including under-the-bark brood size, emergence density (per unit tree surface area), beetle body size, and sex ratio. Large beetles have higher fecundity and can fly farther \citep{safranyik2006biology, graf2012association, evenden2014factors}. The sex ratio is relevant because a higher proportion of females suggests some degree of overwintering stress, since females are the more robust sex; this is evidenced by a higher proportion of males emerging from trees with thinner phloem \citep{cole1976mathematical}.

Statistical models are essential for assessing the effective attack rate in jack pine, particularly because field experiments for this purpose are exceptionally work-intensive. Although researchers can induce mass attacks on bolts (see the previous paragraph), experimentally inducing the entire progression of beetle attacks (including host-finding, host-selection, and mass-attacks) on living trees is much more laborious. Models remain our primary tool for assessing jack pine's impact at the landscape level--- an area where research is critically lacking \citep{bleiker2019risk}. To date, the only modeling study that utilized pine species as a covariate found a strong negative association between MPB infestations and jack pine \citep{srivastava2023dynamic}. 


Here, we take an integrative approach, utilizing both models and bolt experiments to assess the effective attack rate and effective brood size in jack pine. First, we analyze a semi-mechanistic model of MPB dynamics, parameterized with extensive survey data from the government of Alberta. The effective attack rate is low in jack pine, but the effective brood size cannot be precisely estimated due to the small number of infestations in jack pine. However, bolt experiments indicate that the effective brood sizes are similar in lodgepole and jack pine. Next, we analyze a second model in which the effective attack rate depends on pine ancestry and pine volume. Pine ancestry is found to have a larger effect on spread rate, indicating that MPB is responding to some quality of the jack pine phenotype, rather than just the forest structure of eastern Alberta. Finally, using on circumstantial evidence, we propose biological mechanisms that could explain these observed patterns.

\section{Methods} \label{Methods}

\subsection{Data}

The severity and locations of MPB infestations were determined with Heli-GPS surveys and ground surveys. If a tree is killed by MPB, its needles will turn rust-red in the summer or autumn of the following year. The Alberta Ministry of Forestry and Parks uses helicopters to find these \textit{red-topped trees}, recording both the number of trees and GPS coordinates. Field crews go to the locations of red-topped trees and search for nearby \textit{green-attack trees}, trees that are infested but not yet displaying red needles. These trees are ``sanitized'' (i.e., burned or chipped) to prevent the proliferation and spread of MPB. Heli-GPS surveys are a high-quality data source, accurately counting the number of infestations (within $\pm 10$ trees) for 92\% of infestation clusters \citep{nelson2006large} and a positional accuracy of $\pm 30$ meters \citep{government2016mountain}. Ground surveys are even more accurate, with a 98.5\% detection rate within the surveyed area \citep{bleiker2019risk}. 

Our models utilize several other data sources. Estimates of pine volume in the Extended Alberta Vegetation Inventory (AVIE; \citealp{alberta_vegetation_inventory_2022}) were obtained through data-sharing agreements with forestry companies. Predicted pine ancestry values were obtained from \citet{cullingham2012characterizing}. The authors used a genetic microchip to analyze pine DNA samples and processed this data through the Structure software program \citep{pritchard2000inference}. The software calculated Q-values, on a scale from $Q = 0$ (pure jack pine) to $Q=1$ (pure lodgepole pine). Cullingham et al., then used used environmental variables and logistic regression to predict Q-values across Alberta. Throughout the paper we designate jack pine forest as $Q < 0.1$, hybrid forest as $0.1 \leq Q < 0.9$, and lodgepole forest as $Q \geq 0.9$. Lastly, estimates of MPB's overwintering survival are calculated using the model of \citet{regniere2007modeling} as implemented in the \textit{BioSIM} program \citep{regniere2014biosim}.

\subsection*{Data preparation} \label{data_prep}

The raw Heli-GPS and ground survey data contained GPS coordinates for red-topped or sanitized green-attack trees respectively. We rasterized this point data into 1$\times$1 km pixels. Much finer resolutions would have been computationally prohibitive given the extensive size of our study area.

The total number of infested trees in pixel $x$ and year $t$, denoted $I_t(x)$, was calculated as 

\begin{equation}
I_t(x) = m_{t}(x) + r_{t+1}(x),
\end{equation}

where $m_{t}(x)$ was the number of sanitized green-attack trees in the focal year, and $r_{t+1}(x)$ is the number of red-topped trees observed in the following year. Our overarching goal was to predict the location and severity of next year's infestations, given this year's infestations. while $I_t(x)$ is a suitable response variable (i.e., that which is predicted), it is not a suitable predictor (i.e., a model input), because it contains sanitized trees that cannot contribute to future infestations. Therefore, we defined a slightly modified input variable: $I_{t}^{*}(x) = I_t(x) - m_t(x)$.

Several steps were taken to prepare the data for model-fitting. Pixels representing pine volume less than 1 $m^3 km^{-2}$ were removed, as infestations are highly unlikely in these locations; in the model, any beetles dispersing to these pixels are assumed to die. To mitigate the risk of overly narrow credible intervals caused by conditionally non-independent data (i.e., spatially autocorrelated residuals), we thinned the dataset at a 3 km scale by selecting only the intersection of every third column and row of the data rasters. This decision was informed by patterns of spatially autocorrelated residuals from a model that was fitted with all pixels (see Appendix \ref{autocorr}). We tested an alternative approach using a Gaussian process latent variable to handle the spatial autocorrelation. This gave similar results as the spatial thinning, but required 5$\times$5 km pixels to remain computationally tractable. We therefore chose to proceed with the spatial thinning approach for our final analysis (see Appendix \ref{GP} for details).

\begin{figure}[H]
\centering
\includegraphics[scale = 1]{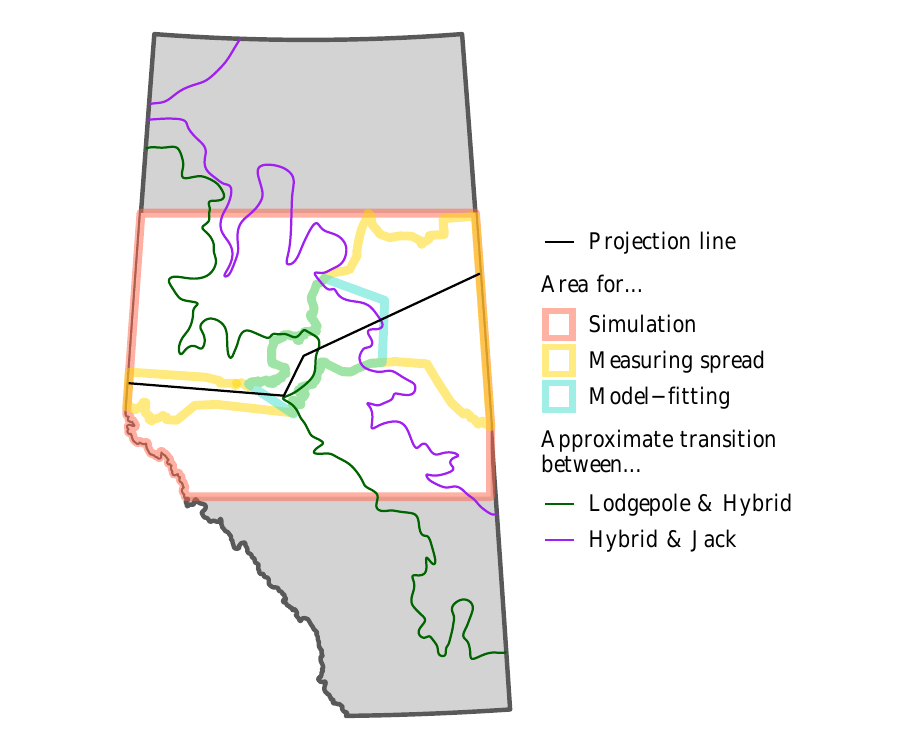}
\caption{Different regions of Alberta are used for different parts of the analysis.}
\label{fig:study_areas}
\end{figure}

Using data from 2006-2015, the model forecasts infestations for 2007-2016. It does not attempt to predict 2006, because many of the 2006 infestations are thought to be caused by long-distance dispersal from British Columbia \citep{carroll2017assessing, bartell2008microsatellite, samarasekera2012spatial}. Lastly, the study focused on a specific region of Alberta with approximately equal lodgepole-dominant and hybrid/jack-dominant pixels (Fig. \ref{fig:study_areas}).

\subsection{Model \#1} \label{mod1}

We developed a statistical model to discern the differences between lodgepole pine and jack pine, with respect to their effective attack rate (i.e., how easy it is to find a suitable host, aggregate, and kill a tree) and effective brood size (i.e., average offspring per tree, multiplied by the average dispersal-phase survival). The model aims to predict the number of trees infested with MPB, based on last year's infestations and the severity of the interceding winter. 

First, we simulated MPB dispersal by convolving last year's infestations with a dispersal kernel. The result was \textit{beetle pressure} \( B_t(y) \), which is proportional to the expected number of beetles arriving at the destination pixel $y$:
\begin{equation} \label{eq:convolve}
B_t(y) = \sum_{x} I_{t-1}^*(x) c(x) \theta_t(x) \bar{D}\left(\text{dist}(y, x)\right) 
\end{equation}

Here, $I_{t-1}^*(x)$ denotes the previous year's infestations that did not end up being sanitized; the parameter $c(x)$ denotes the effective brood size per infested tree, which varies with the species of pine present; and $\theta_t(x)$ denotes the probability of winter survival, pre-computed using the model of \citet{regniere2007modeling}. The quantity $I_{t-1}^*(x) \, c(x) \, \theta_t(x)$ is therefore proportional to the number of beetles produced in pixel $x$. The function $\bar{D}$ is the probability of dispersing from pixel $x$ to pixel $y$. The sum over all pixels represents the discretization of a continuous-space convolution using a midpoint Riemannian sum.

The underlying dispersal kernel is a Student's t-distribution, which allows for a predominance of short-distance dispersal and occasional long-distance dispersal. The kernel density $D$ is radially symmetric in 2D space, and therefore may be parameterized as a function of the Euclidean distance between two pixels, denoted $r = \text{dist}(x,y)$. The non-marginal kernel density is given by 
\begin{equation} \label{convolve2}
D(r) = \frac{(\nu -1) \left(\frac{r^2}{\nu  \rho ^2}+1\right)^{\frac{1}{2} (-\nu -1)}}{2 \pi  \nu  \rho ^2}.
\end{equation}

Based on a previous study of MPB dispersal in western Alberta, the dispersal parameters were fixed at $\nu = 1.45$ and $\rho = 0.0118$ \citep{johnson2024stratified}. To accurately represent dispersal at a 1km resolution, we convolve higher-resolution grids (50x50 m pixels) of the discretized kernel density (i.e., $D(r) \times \left(0.05\right)^2$) and infestations that are uniformly distributed within the central square kilometer. The results of the convolution are aggregated to the 1km scale, resulting in the transition probability function, $\bar{D}$ (see \eqref{eq:convolve}).

The effective brood size $c(x)$ varies based on the type of pine:
\begin{equation}
c(x) = \begin{cases} 
c_L & \text{if } x \in L, \\
c_J & \text{if } x \in J,
\end{cases}
\end{equation}
where \( L \) and \( J \) are the sets of all locations with lodgepole pine and jack/hybrid pine, respectively. The combination of two effective brood size parameters, along with all effective attack rate parameters (to be introduced shortly) are jointly unidentifiable. Therefore, we fixed the lodgepole pine brood size at $c_L = 1$, and allowed $c_J$ to vary; this is precisely why we described $B_t(x)$ as \textit{proportional} to the number of beetles. Only infested trees are observed, so the true number of beetles cannot be estimated from the survey data. 

The ``lodgepole pine pixels'' \( L \) contained both lodgepole pine ($Q > 0.9$) and hybrids ($0.1 \leq Q < 0.9)$. This grouping is the result of model selection --- we tried grouping hybrid pixels with either jack or lodgepole pine pixels, and the latter resulted in better predictions according to leave one out cross validation (see Appendix \ref{robust1}). While the effective brood size could theoretically be a continuous function of Q, this would require computationally expensive convolution at each iteration of the model-fitting procedure. Discretization allowed us to use fast Fourier transforms for dispersal calculations just once before model fitting begins. Then, we simply multiplied the \( J \) dispersal surface by $c_J$. 

To predict next year's infestations from beetle pressure, we adopted a \textit{zero-inflated negative binomial} model. This choice naturally accommodates two key features of the data: 1) a high proportion of sites with no infestations (i.e., zero-inflation), and 2) overdispersion in the count data (i.e., the variance exceeds the mean number of infestations). The high proportion of zeros likely reflects high levels of dispersal mortality \citep{safranyik1989mountain, latty2007pioneer, pope1980allocation} and within-pixel aggregation. Zero-inflated negative binomial models are common in ecology and have previously been used to model MPB \citep{xie2024modelling}.

The partial probability of observing an infestation, denoted $\pi_t(x)$, was modeled using logistic regression. This presence/absence sub-model only generates the \textit{potential} for a non-zero number of infestations, as the count model may still generate zeros.

\begin{equation} \label{eq:pi}
\pi_t(x) = \begin{cases} 
\text{logit}^{-1}\left(\gamma_{0,L} + \gamma_{1,L} \log\left(B_t(x)\right) \right) & \text{if } x \in L, \\ 
\text{logit}^{-1}\left(\gamma_{0,J} + \gamma_{1,J} \log\left(B_t(x)\right) \right) & \text{if } x \in J.
\end{cases}
\end{equation}

The number of infestations follows a negative binomial distribution, parameterized by the mean ($\mu$) and the dispersion parameter ($\phi$):
\begin{equation} \label{eq:count}
f(I_t(x)) = \begin{cases} 
\text{NB}\left(I_t(x) \; | \; \mu = \exp\left[\beta_{0,L} + \beta_{1,L} \log\left(B_t(x)\right)\right], \phi = k_L \right) & \text{if } x \in L, \\
\text{NB}\left(I_t(x) \; | \; \mu = \exp\left[\beta_{0,J} + \beta_{1,J} \log\left(B_t(x)\right)\right], \phi = k_J \right) & \text{if } x \in J.
\end{cases}
\end{equation}
Here, $\text{NB}(y \; | \; \mu, \phi)$ denotes the probability mass function, 
\begin{equation}
\text{NB}(y \; | \; \mu, \phi) = \binom{y + \phi - 1}{y} \left(\frac{\mu}{\mu + \phi}\right)^y \left(\frac{\phi}{\mu + \phi}\right)^\phi
\end{equation}

The overall probability of observing a number of infestations is a mixture of the zero-inflated and the negative binomial components:
\begin{equation}
\text{Pr}(I_t(x)) = \begin{cases} 
(1-\pi_t(x)) + \pi_t(x) \cdot f(I_{t}(x)) & \text{if } I_t(x) = 0, \\
\pi_t(x) \cdot f(I_{t}(x)) & \text{if } I_t(x) > 0.
\end{cases}
\end{equation}

The structure of the statistical model is justified by graphical evidence and prior knowledge. The logistic relationship in the presence/absence model and the log-log linear relationship in the count model emerge from exploratory plots (Fig. \ref{fig:mod1_just}), and the negative binomial distribution accurately accounts for residual error (Fig. \ref{fig:NB_pred_vs_obs}). The model structure is also biologically sensible. Mountain pine beetle experiences positive density dependence in the form of an Allee effect, in that a threshold number of beetle attacks are required to overcome tree defenses \citep{thalenhorst1958grundzuge, raffa1983role, boone2011efficacy}. Mountain pine beetle also experiences negative density dependence at larger spatial scales, putatively due to the difficulty of finding suitable host trees that are not already filled up with beetles \citep{trzcinski2009intrinsic, macquarrie2011density}. Both kinds of density dependence occur at the host-finding/selection and mass-attack stages of MPB's life cycle, and are therefore plausibly captured by the \textit{effective attack rate parameters}, e.g.,  $\beta_{0,L}, \beta_{1,L}, \gamma_{0,L},$ and $\gamma_{1,L}$ for lodgepole pine alone.

The \textit{effective brood size} is represented by a single parameter (e.g., $c_L$ for lodgepole pine) because it appears to be independent of beetle pressure, as shown in the \textit{Alberta disk data} (Appendix \ref{disk}). This consistency in brood size is likely due to the mountain pine beetle (MPB) maintaining a near-optimal attack density on successfully attacked trees (approximately 60 attacks per square meter; \citealp{raffa1983role}), regardless of overall beetle population pressure. Beetles achieve this density through two main mechanisms: anti-aggregation pheromones, which direct excess beetles to nearby trees, and male stridulations (sound signals), which help space galleries evenly \citep{safranyik2006biology}. Beetles ignore these warning signals, possibly due to interference competition with established mating pairs and the poor quality (i.e., thin phloem) of the remaining uncolonized areas \citep{raffa2001mixed}.

\subsection{Model \#2} \label{mod2}

We created a second model to determine whether MPB is more influenced by pine species identity or structural differences between the forests of western and eastern Alberta. Pine volume served as a proxy for forest structure, capturing the west-to-east gradients in tree size, tree density, forest composition, and pine connectivity. The model generally has the same structure as Model \#1 --- a dispersal step to calculate beetle pressure, followed by a zero-inflated negative binomial distribution --- with a few notable differences. First, we fixed both brood size parameters at $c_L = c_J = 1$, which was based on the fact that $c_J$ could not be estimated precisely by model \#1, and the fact that prior bolt experiments indicate similar brood sizes (see the \textit{Results}). Second, instead of allowing for pine-specific attack rate parameters (e.g. $\gamma_J$ \& $\gamma_J$), we use both pine ancestry and pine volumes as predictor variables.

The probability of potentially observing a non-zero number of infestations (previously \eqref{eq:pi}), now becomes
\begin{equation}
\text{logit}\left(\pi\right) = 
\gamma_{0} +  \gamma_{B}  B^{*}  + \gamma_{Q} Q^* + \gamma_{V}  V + \gamma_{QV} Q  V,
\end{equation}
where $B^*$ $Q^*$ and $V$ are respectively the beetle pressure, pine ancestry, and pine volume predictors. Dropping the explicit notation for time and location, and defining the standardization operator $\text{std}\left((x-\overline{x})/\text{sd}(x)\right)$, we can write the predictors as  as $B^* = \text{std}\left(\log(B)\right)$, $Q^* = \text{std}\left(Q\right)$, $V = \text{std}\left(\log(\text{pine vol.})\right)$. The log-transformations were necessary to meet linearity assumptions, and standardization allowed us to interpret the magnitude of the coefficients as predictor importance.





A similar modification was made to the count model (previously \eqref{eq:count}); the mean of the negative binomial distribution becomes
\begin{equation}
\mu = \exp\left[\beta_{0} +  \beta_{B}  B^{*}  + \beta_{Q} Q^* + \beta_{V}  V + \beta_{QV} Q V \right].
\end{equation}
Again, the structure of model \#2 is justified by graphical evidence: the predictors $B^*$, $Q^*$, and $V$ all have approximately linear relationships with $\text{logit}\left(\pi\right)$ and $\log\left(\mu\right)$ (Fig.\ref{fig:mod2_just}). The types of model inputs used across both model \#1 and model \#2 --- beetle densities, winter temperatures, pine species, and pine volume/density --- are consistently found as the most important predictors of MPB dynamics \citep{aukema2008movement, ramazi2021predicting, srivastava2023dynamic}. As a robustness check, a variant of model \#2 with interaction effects is analyzed in Appendix \ref{mod2_int}.

Both model \#1 and model \# 2 were parameterized using \textit{Stan}, a Bayesian model-fitting program. Standard model-fitting diagnostics were examined. Additionally, the posterior contraction statistic demonstrated that prior distributions generally had a small influence on parameter estimates (Appendix \ref{model_details}).

\subsection{Simulations} \label{simulations}

We simulated both models across the entire longitudinal extent of Alberta (Fig. \ref{fig:study_areas}), using the 2006 infestations as the initial state (Fig. \ref{fig:sim_spread_w_inset}). Infestations were not allowed in pixels where pine volume was less than 1 $\text{m}^3 \text{km}^{-2}$. To simulate the provincial control efforts, we define $m_t$, the proportion of infestations that were controlled in year $t$ (within the polygon for measuring spread; Fig. \ref{fig:study_areas}). Then, the calculation of beetle pressure (previously \eqref{eq:convolve}) becomes
\begin{equation} \label{eq:convolve2}
B_t(y) = \sum_{x} I_{t-1}(x) m_{t-1} c(x) \theta_t(x) \bar{D}\left(\text{dist}(y, x)\right), 
\end{equation}
where $I_{t-1}(x) \times m_{t-1}$ is the simulated number of uncontrolled infestations. 

To measure MPB spread, we took all data within the \textit{Area for measuring spread} (Fig. \ref{fig:study_areas}) and projected it onto a \textit{projection line} with runs west-to-east across Alberta. Then, eastward spread was measured as the $99^{\text{th}}$ percentile of eastward distance along this line. The \textit{Area for measuring spread} polygon is the intersection of surveyed areas across years, and thus allows for a fair comparison between real and simulated data; the projection line was intentionally drawn to transverse through the middle of this polygon.

The simulations served several purposes. First, a comparison of actual data and simulated data helped validate the models. Second, simulations quantified the uncertainty in the eastward spread of MPB infestations. Each simulation was instantiated with a random sample from the joint posterior distribution of model parameters, and thus variation across simulations represents both parameter uncertainty and accumulated process error. Third, We performed of variety of such counterfactual simulations for both model \#1 and model \#2, respectively focused on teasing out the relative importance of the effective attack rate versus the effective brood size, and pine ancestry versus pine volume.


The statistical models did not account for the depletion of suitable pine hosts, but we included this factor in the simulations. We imposed a limit on the number of trees that can be infested in a single pixel, as otherwise, an exponential buildup in some locations could create unrealistic beetle pressure. The maximum number of trees that can be infested per pixel, across all years, is set to the $99^{\text{th}}$ percentile of cumulative infestations across all pixels containing lodgepole pine (i.e., $Q > 0.9$) and at least one infestation. Resource depletion is not included in the statistical model for pragmatic reasons, and because it is unclear whether the effects of resource depletion are estimable --- MPB populations have declined across Alberta, but this may be primarily attributable to the provincial government's control efforts or several unusually cold winters from 2018--2023, rather than host tree depletion via beetle attack.

\section{Results} \label{Results}
Our results indicate that MPB’s slow spread through jack pine is attributable to a low effective attack rate, not a low effective brood size. The first statistical model was not able to accurately estimate the effective brood size, but experimental studies strongly suggest that the two pine species have similar effective brood sizes \citep{safranyik1982survival, cerezke1995egg, rosenberger2017cold, musso2023pine}. The second statistical model implies that the low effective attack rate in jack pine is due to jack pine's phenotype \textit{and} and the lower pine volumes of eastern Alberta; however, pine species identity is the more important predictor.

The effective brood size cannot be discerned using Model \#1. Specifically, the effective brood size in jack pine is estimated with wide credible intervals, predicting 30-160\% of the effective brood size in lodgepole pine (Table. \ref{tab:pars1}). The reason for this large uncertainty is simple: there are so many more infestations in lodgepole pine that the total beetle pressure is dominated by contributions from lodgepole pine stands, regardless of the jack pine's brood size multiplier (i.e., the parameter $c_J$).

To probe the robustness of this result --- the ineffectiveness of models in estimating the effective brood size --- we examined the posterior distribution of $c_J$ across a number of different subjective modeling decisions (Appendix \ref{robust1}). While some models expressed confidence (i.e., narrow posteriors), they disagreed on whether jack pine or lodgepole pine had the larger effective brood size. This disagreement further demonstrates that models are not effective for estimating  $c_J$.  

\begin{figure}[H]
\centering
\includegraphics[scale = 1]{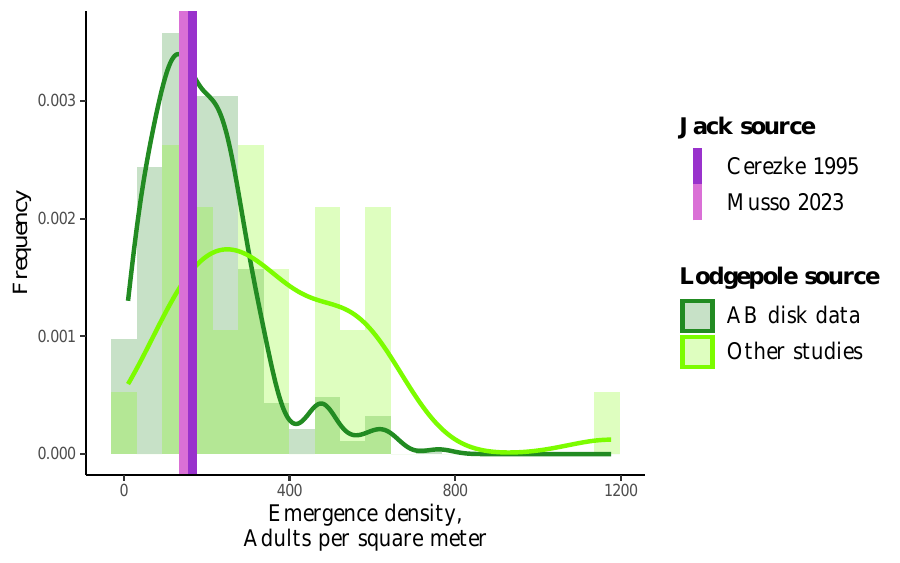}
\caption{Emergence densities are similar in jack pine and lodgepole pine. The studies assessing emergence density in jack pine are \citet{cerezke1995egg} and \citet{musso2023pine}. ``AB disk data'' contains emergence densities from thousands of lodgepole pine across Alberta \citep{government2016mountain}. ``Other studies'' contains assorted estimated, often from United States Forest Service reports: \citet{tishmack2005mountain, negron2018biological, reid1963biology, raffa1983role, rasmussen1980emergence, de1939biology, safranyik1988estimating, cole1969mountain, schmid1972emergence, whiteside1937progress, beal1938progress, peterman1974some, klein1978attack, safranyik1985relationship,safranyik1991unseasonably}. }
\label{fig:emerge_hist}
\end{figure}

Bolt experiments show that the effective brood size is similar in lodgepole pine and jack pine (Table \ref{tab:bolt}). Of the five experiments that have been conducted in both pine species, four studies find similar MPB fitness. The exceptional study \citep{bleiker2023suitability} only used a single tree per pine species, so the result may be due to a tree effect, rather than species effect. Of the two experiments that measured emergence density (progeny per tree surface area), both found similar emergence densities in lodgepole pine and jack pine \citep{cerezke1995egg, bleiker2023suitability}. Further, experimental emergence densities in jack pine bolts are similar to the median emergence density in lodgepole pine in Alberta, per the \textit{Alberta disk data} (Fig \ref{fig:emerge_hist}). Finally, beetle body size and sex ratios are generally similar in lodgepole and jack pine (Table \ref{tab:bolt}).

\renewcommand*\rot{\multicolumn{1}{R{50}{1em}}}
\begin{landscape}
\begin{table}[h!] 
\centering
\caption{The effective brood size of MPB is similar in lodgepole pine and jack pine: evidence from bolt experiments. Values are given as mean $\pm$ standard error (if available). Blank cells represent N/A values. Only two studies \citep{cerezke1995egg, musso2023pine} attempted to simulate mass attack; these studies resulted in lower fitness measurements, putatively due to higher intraspecific competition between larvae. The remaining studies inoculated mating pairs into pre-drilled holes. One study was considered but excluded due to the presence of saprophytic fungi in some of the bolts used \citep{lusebrink2016effect}. \\
\small{
Notes:\\
(1) Prothorax width, not pronotum width.\\
(2) emerging adults per female, not under-bark brood per female.\\ 
(3) DBH range is across lodgepole and jack pine bolts together. \\
(4) Midwinter (January) brood as opposed to spring/summer pre-flight brood.} }
\small
\begin{tabular}{llllllllllll}
\rot{Data source} & \rot{Pine species} & \rot{\makecell[l]{Fitness proxy,\\brood/female}} & \rot{\makecell[l]{Emergence density,\\adults per $\text{m}^2$}} & \rot{\makecell[l]{Pronotum width,\\female/male, in mm}} & \rot{\makecell[l]{Proportion female}} & \rot{\makecell[l]{No. mating pairs introduced}} & \rot{No. trees} & \rot{No. bolts} & \rot{Phloem thickness, mm} & \rot{DBH, cm} & \rot{Successful galleries} \\
\hline
\citet{safranyik1982survival} & Lodgepole & 31.6 &  & 2.12/$1.89^{(1)}$ & 0.64 &  &  &  &  &  &  \\
 & Jack & 36.5 &  & 2.09 (0.05) / $1.90 (0.032) ^{(1)}$ & 0.81 & 20 & 1 & 4 &  & 19.9 & 65\% \\
\hline 
\citet{cerezke1995egg} & Lodgepole & $3.94^{(2)}$ & 129 & 2.13 / 1.91 & 0.625 &  & 2 & 6 &  & 19--$23^{(3)}$ &  \\
 & Jack & $2.59^{(2)}$ & 165 & 1.99 / 1.79 & 0.513 &  & 2 & 6 &  & 19--$23^{(3)}$ &  \\
\hline
\citet{rosenberger2017cold} & Lodgepole & 22.9 $(3.2)^{(4)}$ &  &  & 0.553 &  & 8 & 8 &  & 23--$30^{(3)}$ &  \\
 & Jack & 27.8 $(4.2)^{(4)}$ &  &  & 0.549 &  & 8 & 8 &  & 23--$30^{(3)}$ &  \\
\hline 
\text{\citet{bleiker2023suitability}} & Lodgepole & 38.4 (6.1) &  & 1.97 (0.02) / 1.76 (0.02) & 0.5 & 26 & 1 & 1 & 1.3 & 30--$40^{(3)}$ & 54\% \\
 & Jack & 15.1 (5.0) &  & 2.01 (0.01) / 1.82 (0.02) & 0.625 & 35 & 1 & 1 & 1.6 & 30--$40^{(3)}$ & 49\% \\
\hline 
\makecell[l]{\citet{musso2023pine},\\ch. 2} & Lodgepole & 2.35 $(0.976)^{(2)}$ & 139 (5.1) & 1.99 (0.086) & 0.673 & 14 & 28 & 13 & 1.99 (0.086) & 26.2 (0.48) &  \\
\makecell[l]{\citet{musso2023pine},\\ch. 3} & Jack & 3.82 $(0.558)^{(2)}$ & 143 (36.8) & 1.4 (0.056) & 0.588 & 18 & 36 & 13 & 1.4 (0.056) & 29.0 (0.54) &  \\
\hline
\end{tabular}
\label{tab:bolt}
\end{table}
\end{landscape}

\begin{figure}[H]
\centering
\includegraphics[scale = 1]{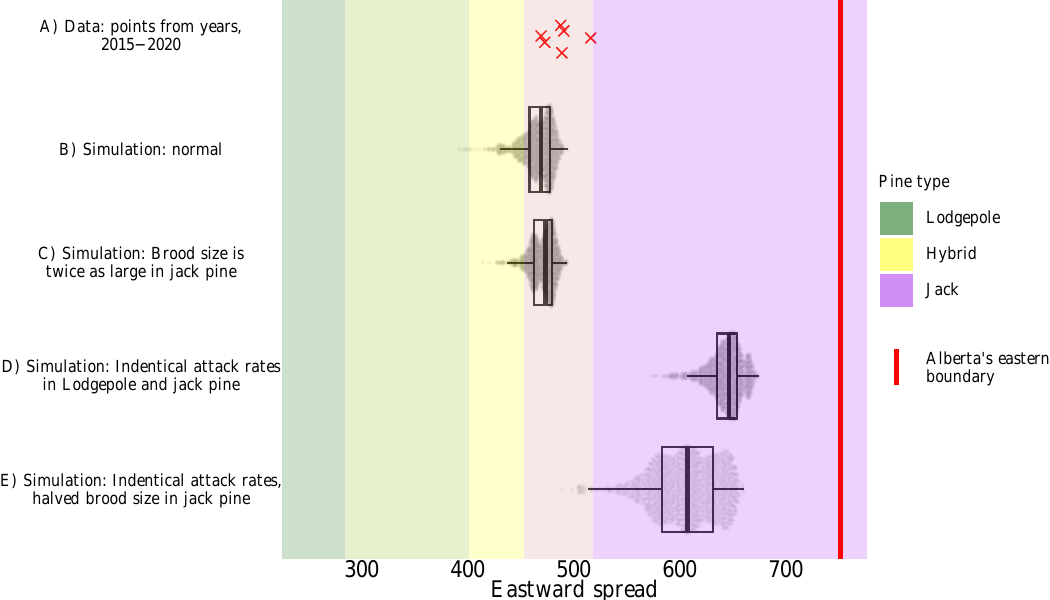}
\caption{Mountain pine beetle's slow spread is due to a low attack rate in jack pine: evidence from counterfactual simulations of model \#1. The y-axis shows demarcates the data and different simulation scenarios. The x-axis shows eastward spread as measured by the 99th percentile of infestation distance, from 2015--2020, along the projection line (see Fig \ref{fig:study_areas}). }
\label{fig:sim_boxplot}
\end{figure}

Counterfactual simulations of Model \#1 imply that a low effective attack rate is responsible for MPB’s slow spread through jack pine. When we change the attack rate parameters in jack pine so that they are equal to the corresponding attack rate parameter in lodgepole pine, MPB moves quickly through eastern Alberta (Fig. \ref{fig:sim_boxplot}, row E). In contrast, increasing the effective brood size to 2 $\times$ larger in jack pine (compared to lodgepole pine) does not substantially increase the spread distance (Fig. \ref{fig:sim_boxplot}, row C). This indicates that MPB has such a low attack rate in jack pine forests, that even a large brood size cannot compensate for it. If the effective brood size to 2 $\times$ \textit{smaller} in jack pine, but MPB has identical attack rates in both pine species, then MPB still spreads through Alberta (Fig. \ref{fig:sim_boxplot}, row D). Rows C and D respectively show that having similar effective brood sizes in lodgepole and jack pine is neither sufficient nor necessary for the spatial spread of MPB through jack pine.

\begin{figure}[H]
\centering
\includegraphics[scale = 1]{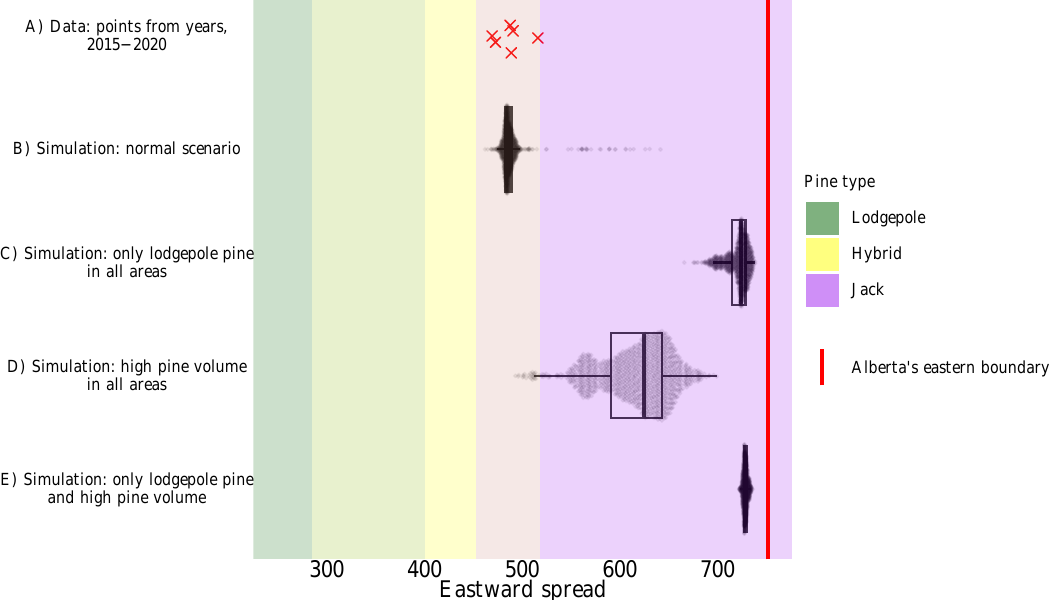}
\caption{Mountain pine beetle's slow spread is primarily due to some species-level property of jack pine: evidence from counterfactual simulations model \#2. The y-axis shows demarcates the data and different simulation scenarios. The x-axis shows eastward spread as measured by the 99th percentile of infestation distance, from 2015--2020, along the projection line (see Fig \ref{fig:study_areas}). }
\label{fig:sim_vol_boxplot}
\end{figure}

Counterfactual simulations of Model \#2 imply that jack pine's phenotype is responsible for MPB’s slow spread through eastern Alberta. When we model a scenario where pine is everywhere –-- setting the pine ancestry predictor to $Q = 1$ throughout the simulation area –-- MPB spreads further eastward in comparison to the baseline simulations (Fig. \ref{fig:sim_vol_boxplot}, row C). To investigate the effect of low pine volumes in eastern Alberta, we imagine that pine volume in western Alberta is replicated everywhere --- all pixels are assigned the mean pine volume of lodgepole pine forests (Fig. \ref{fig:sim_vol_boxplot}, row D); the mean is calculated across pixels within the area for measuring spread (see Fig. \ref{fig:study_areas}) where $Q > 0.9$, and pine volume $> 1 \text{m}^3 \text{km}^{-2}$. Here, we see a spread distance that is greater than the normal normal scenario, but less than the ``lodgepole everywhere'' scenario. This suggests that both pine species and pine volume matter, but that the pine species identity matters more.

\section{Discussion} \label{Discussion}

The slow spread of mountain pine beetle (MPB) through eastern Alberta is due to an inability to find and attack jack pine, which in turn is largely caused by some species-level properties of jack pine. Importantly, our results contrast with a commonly accepted view, wherein the slow spread of MPB is attributed to low effective brood sizes (i.e., poor reproduction and development) and lower pine volumes.  For instance, a recent national (Canadian) risk assessment concluded that “The slower rate of spread is attributed to lower pine volumes, poor connectivity of susceptible pine stands, lower probability of long-distance dispersal events as distance to a large source population increases, and aggressive control efforts sustained to date by the province in eastern Alberta” \citep{bleiker2023suitability}. An earlier paper had a similar conclusion: ``Jack pine stands in the boreal zone are less susceptible to outbreaks than are lodgepole pine stands in western Canada because of relatively lower pine volume in these stands.'' \citep{safranyik2010potential}. In contrast, we find that landscape-scale differences in pine volume play a secondary role in the eastward spread of MPB.

A variety of mechanisms can be invoked to explain MPB’s differential response to lodgepole versus jack pine. In the following paragraphs, we discuss mechanisms as they apply to different stages of the MPB life cycle. An executive summary with subjective certainty ratings is provided in the \textit{Conclusions}.


\subsection*{MPB development and reproduction}

Previous studies have identified several mechanisms affecting MPB development in jack and lodgepole pine, including differences in phloem thickness and nitrogen content. Jack pine tends to have thinner phloem than lodgepole pine due to its smaller diameters and lower phloem thickness at the same diameters (\citealp{cole1973estimation, lusebrink2016effect}; Fig. \ref{fig:phloem_dbh}). Since MPB larvae consume phloem as their primary source of nutrition, thin phloem could plausibly lead to slower development, more overwintering mortality, smaller body sizes, and less fat content \citep{cole1973estimation, musso2023naive}. Jack pine phloem has 30\% lower nitrogen content, potentially impacting larval growth and development \citep{lusebrink2016effect}.

While thinner phloem and lower nitrogen content in jack pine imply slower MPB development and higher overwintering mortality in theory, these factors seem to have a minor impact in reality, putatively because of compensatory behaviors observed in MPB larvae. \citet[Ch. 3]{musso2023pine} found that larvae in jack pine compensate for thinner phloem by consuming across a greater area of the bole; the total volume of phloem consumed is actually greater in jack pine than in lodgepole pine.  Jack pine is poorly defended, with lower levels of constitutive defensive compounds compared to lodgepole pine \citep{clark2014comparison}, but this does not seem to have a net-effect on MPB fitness. Perhaps reduced mortality from resinosis is counteracted by intraspecific competition between larvae \citep{raffa1983role}, or brood mortality from resinosis is simply not a large component of total mortality \citep{amman1983mountain}.


Jack pine has also produces lower levels of \textit{induced} defensive compounds in response to MPB attacks \citep{musso2023naive} and inoculation with MPB's symbiotic blue-stain fungus \citep{arango2016differences, lusebrink2016effect, erbilgin2017water}, which should reduce the larval mortality due to toxicity. We can infer this effect by drawing parallels with lodgepole pine: While trees in the historic range exhibit increased terpene concentrations six weeks post-attack \citep{clark2012legacy}, trees in the expanded range show no such increase \citep{musso2023naive}. This lack of induced defenses may explain the higher MPB fitness that has been observed in naïve lodgepole pine \citep{cudmore2010climate}. However, the real-world impact of jack pine's lack of induced defenses is unclear, as experimental methods using cut bolts may not accurately reflect physiological processes within live trees.

Some bolt experiments in Table \ref{tab:bolt} attempt to control for tree size by selecting lodgepole and jack pine bolts within a narrow DBH range. In reality, jack pine trees typically have a 2.5--5 cm smaller DBH than lodgepole pines, on average (\citealp{nunifu2009compatible}, Table 1; \citealp{strimbu2017deterministic}, Table 1; \citealp{burns1990silvics}, p. 617 \& 566). The smaller average size of jack pine could result in a slightly reduced effective brood size compared to lodgepole pine, but given the small difference in average sizes, and the fact that MPB seems to have positive fitness on smaller (i.e., 20--30 cm DBH) lodgepole pines \citep{johnson2024explaining}, this possible difference is likely not large. Additionally, the effect of jack pine's lower induced defenses, which is similarly not reflected in bolt experiments, may counteract any size-related effects.

\subsection*{Host finding/selection}

Due to its chemical profile, jack pine may be harder to locate during the host-finding/selection stage. Jack pine has lower relative and total levels of $\beta$-phellandrene \citep{miller1990beta, clark2014comparison, burke2016influence}, the only monoterpene known to play a role in primary attraction \citep{miller1990beta}. However, it is unknown how important olfaction is to MPB during the host-finding/selection stage compared to other senses \citep{reid1963biology, raffa1982gustatory, raffa2001mixed}. While lower $\beta$-phellandrene levels might reduce jack pine's attractiveness to MPB, high beetle pressures likely mitigate this effect, since only a few attacking females are needed to initiate a mass attack.

\subsection*{Mass attacks}

During MPB mass attacks, differences in the chemical profiles of lodgepole and jack pine have been hypothesized to influence MPB behavior. Jack pine exhibits higher relative concentrations of $\alpha$-pinene, the precursor to MPB’s primary aggregation pheromone, \textit{trans}-verbenol \citep{hughes1973dendroctonus, chiu2018monoterpenyl}, and female MPB colonizing jack pine in the lab produce 2$\times$ more \textit{trans}-verbenol \citep{erbilgin2014chemical}. Jack pine also has lower relative concentrations of myrcene and higher levels of 3-carene, terpenoid volatiles that act synergistically with aggregation pheromones to attract beetles \citep{erbilgin2014chemical, seybold2006pine, chiu2022mountain}. Among these interspecific differences, the most biologically significant is thought to be $\alpha$-pinene, due to its status as a precursor to aggregation pheromones, and due to the fact that its relative concentration is 3--4 higher in jack pine than in lodgepole pine \citep{burke2016influence, erbilgin2014chemical}.

Despite an abundance of $\alpha$-pinene in jack pine, we speculate that jack pine is still difficult to mass attack due to lower total concentration of other crucial monoterpenes.  While there is some experimental evidence that MPB behavior is more influenced by the relative concentration/composition of monoterpenes (reviewed in \citealp{erbilgin2019phytochemicals}), the reliance on bolts or pheromone traps in these studies raises questions about the generalizability of the findings to the real world. In contrast, lodgepole trees that are naturally attacked during an MPB epidemic have nearly twice the total concentration of monoterpenes, in comparison with non-attacked trees \citep{boone2011efficacy}. Across various similar studies involving different bark beetle species and their host trees, total monoterpenes are consistently better predictors of beetle attack than specific monoterpene composition (8 out of 15 vs. 2 out of 15 studies; see \citealp{howe2024quantification}, Table 4). It is likely that total monoterpene concentration, which is typically measured in units of mg/g phloem in the immediate vicinity of boring holes, is a proxy for the chemical environment on a larger spatial scale.

We propose that the physical characteristics of jack pine lead to smaller plumes of attractive volatiles around mass-attacked trees, thus leading to a low effective attack rate. Together, jack pine's smaller size and lower total monoterpenes may result in smaller plumes of attractive volatiles around mass-attacked trees, making it more challenging for MPB to successfully execute mass attacks. It is difficult to say how much of jack pine’s smaller size and thinner phloem  is a species level trait due to the confounding effect of the environment; jack pine tends to grow at lower elevations, with well-drained soil and southern exposure, leading to nutrient poor soil \citep{cullingham2012characterizing, burns1990silvics}.


Supporting evidence for this mechanism --- tree size dependent pheromone/volatile plumes --- comes from extensive studies on lodgepole pine. Invariably, MPB outbreaks end before most medium-sized lodgepole pine are killed, despite beetles having positive fitness when attacking medium-sized trees (\citealp{johnson2024explaining}, Figures 5 \& 6); this implies that MPB need a strong chemical signal (i.e. such as those provided by large trees) to successfully aggregate. Tree diameter is a clear predictor of tree mortality (reviewed by \citealp{bjorklund2009diameter}). Stand-level maps of infestations reveal that beetles only attack medium-sized trees when they are adjacent to large trees \citep{mitchell1991analysis, preisler1993colonization}. Mechanistic models of pheromone diffusion imply that pheromones can be detected 20-50 meters from their source tree, supporting the general idea that MPB have a limited ability to detect host trees \citep{biesinger2000direct, strohm2013pattern}.


Our results show that pine volume is not the best predictor of MPB spread, which speciously contradicts the idea that the lower attack rate on jack pine is partly due to jack pine's smaller size; after all, pine volume can be mathematically represented as the average volume per pine, multiplied by the total number of pines. However, the correlation between pine volume and pine crown cover was $r=0.75$, indicating that pine volume mostly tracks the density and distribution of trees, not tree size. It is whether pine volume across a $\text{km}^2$ pixel is a relevant metric --- from the perspective of MPB, what matters is the availability of dense stands of large-diameter pines. Future studies could attempt to distinguish between the effects of tree size and other species-specific properties (like thinner phloem and lower monoterpene concentrations in jack pine) using stand-level data of tree diameter distributions.

\subsection*{Alternative hypothesis for slowed spread}

There are several alternative explanations for the slow spread of MPB in eastern Alberta, but we do not find evidence for these explanations in our data. One explanation is that the climate of the jack pine forest is unsuitable for MPB. While northern boreal forests are too cold for MPB survival \citep{carroll2004effects, cooke2017predicting}, mid-latitude Alberta jack pine forests have a suitable climate, comparable to lodgepole pine in the Rocky Mountain foothills. This is partially because the lower elevation of eastern Alberta compensates for the higher latitudes where jack pine are found (Fig. \ref{fig:covariate_maps_jack_temp}).

Another explanation attributes the slowed spread to the provincial government's control strategy of cutting and burning infested trees. However, despite control efforts starting in 2006, MPB continued rapid spread until 2010. Moreover, the highest proportion of controlled infested trees was in western Alberta (Fig. \ref{fig:cntrl_1D}). While management actions may have limited total tree mortality \citep{carroll2017assessing} and prevented some eastward spread, they were likely not the primary factor in the significant slowdown around 2009.

A third explanation attributes the rapid eastward expansion to long-distance dispersal from BC, with the slowing spread in Alberta reflecting BC's declining hyperepidemic. A population genetic analysis found that beetles in 2006 and 2007 had different origins compared to those in 2005 \citep{samarasekera2012spatial}, suggesting that the 2006 expansion is at least partially due to long distance dispersal. However, it seems unlikely that subsequent expansions (such as those in 2009 or 2011) were the consequence of long distance dispersal from British Columbia, given the 500 km distance between central BC and central Alberta, as well as decreasing infestation densities in BC.

\section{Conclusions} \label{conclusions}

Our main findings and their interpretations with subjective confidence ratings (low, moderate, and high) are:

\begin{enumerate}

\item MPB’s range expansion has slowed because MPB has difficulty in finding and/or mass-attacking jack pine in the forests of eastern Alberta, not because host trees are inhospitable in terms of  beetle reproduction and development; in the jargon of this paper, MPB has a lower \textit{effective attack rate} in jack pine forests than in lodgepole pine forest (high confidence); but MPB does not have a lower \textit{effective brood size} in jack pine forest (moderate confidence).

\item A modest experimental literature (5 studies) shows that MPB has similar fitness and brood size in jack pine and lodgepole pine bolts. Jack pine has thinner phloem than lodgepole pine, which ought to negatively impact MPB development and survival. However, the majority of bolt experiments (4 out of 5) did not attempt to control for phloem thickness, but still found similar brood sizes in lodgepole and jack pine. It is possible that MPB larvae compensate by simply eating more phloem (low confidence: only a single study measured phloem consumption). It is also possible that the deleterious effects of jack pine's thinner phloem are counteracted by lower levels of constitutive and induced defenses (i.e., toxic monoterpenes) in jack pine (low confidence, bolt experiments cannot assess induced defenses).

\item Pine species identity was a better predictor of MPB spread than pine volume. This is possibly due to the semiochemical profile of jack pine --– specifically lower relative and total concentrations of $\beta$-phellandrene --- making host-finding more difficult (low confidence). The slightly more plausible mechanism, however, involves a combinaion of jack pine’s smaller size, thinner phloem, and lower total monoterpene concentrations; these traits result in smaller plumes of attractive volatiles during the mass-attack stage, thus decreasing the probability of a successful attack (moderate confidence).
\end{enumerate}

\begin{figure}[H]
\centering
\makebox[\textwidth]{\includegraphics[scale = 0.85]{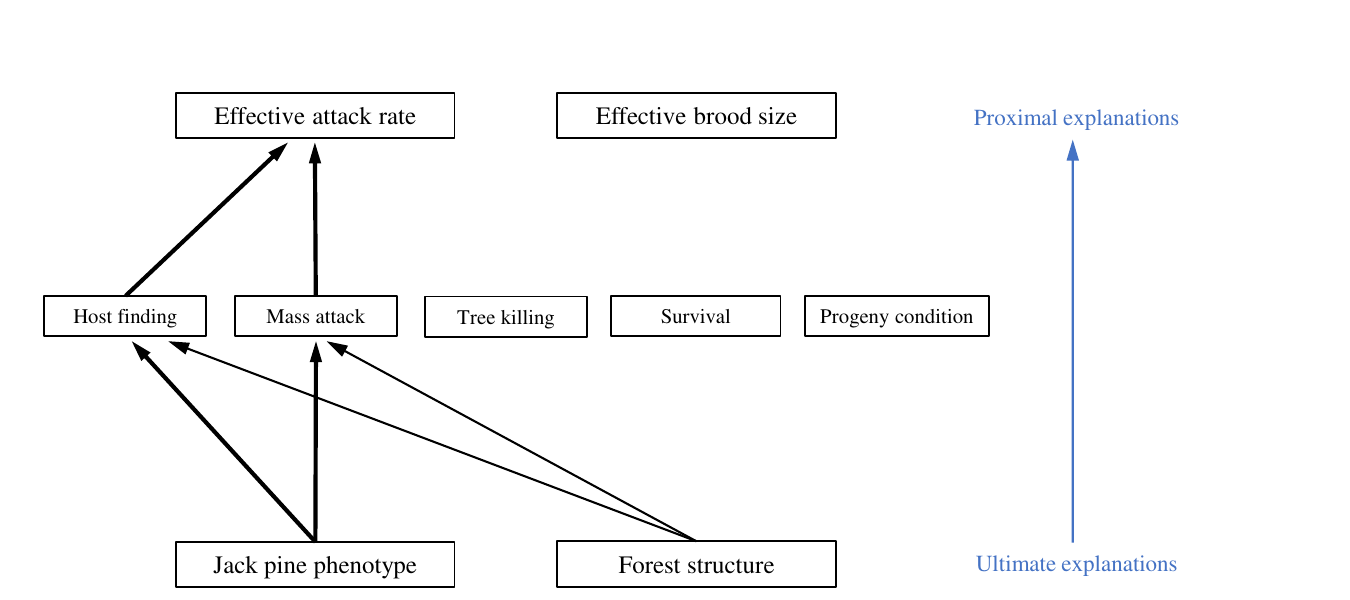}}
\caption{Conclusions, visualized as a simplification of Figure \ref{fig:explanation_hierarchy1}. Some species-level property of jack pine (putatively its smaller size, thinner phloem, and lower total monoterpene concentrations), negatively affects MPB's ability to find and/or mass-attack jack pine tree. The thinner arrow indicate a smaller effect. The mediating arrow through ``Tree killing'' has been removed because it is assumed that there are always sufficient beetles to overcome tree defenses during the epidemic phase of an outbreak.}
\label{fig:explanation_hierarchy2}
\end{figure}

The notion that jack pine is inherently unsuitable for mountain pine beetle (MPB) should be approached cautiously. The most probable scenario for further range expansion is a long-distance dispersal event from high-volume lodgepole pine stands in central Alberta to high-volume jack pine stands in western Saskatchewan \citep{bleiker2019risk, hodge2017strategic} Our results might be misinterpreted to suggest that MPB management in central Alberta is unnecessary, as MPB supposedly won't persist in jack pine. However, Figure \ref{fig:sim_vol_boxplot} clearly shows that pine volume also matters. It is also possible that MPB is responding to the availability of large-diameter trees, and that both pine species identity and large-diameter tree availability could be strongly correlated due to latent environmental variables not reflected in the pine volume predictor. Such a confounding bias is unlikely, but not impossible. 

Our semi-mechanistic models have several additional limitations. First, our models assume that beetle attack rate parameters are only affected by the forest properties in the destination pixel; in reality, beetles probably respond to forest properties along their flight path. Second, since the simulations operate at the $1\text{km}^2$ spatial scale, sub-pixel patchiness in the distribution of pine trees (i.e., forest connectivity) is assumed to not affect MPB dynamics. Third, the model underestimates real-world spatial autocorrelation of infestations (Fig. \ref{fig:sim_map}), possibly due to clustered dispersal from unpredictable weather. As a consequence, the model is not suitable for analyzing medium-scale spatial patterns; e.g., correlation between patches separated by up to 10 km.

Our study shows that MPB's slow spread through eastern Alberta is mainly due to problems finding and attacking jack pine, rather than issues with reproduction or development. While the amount of pine in an area matters, some inherent properties of jack pine (putatively smaller sizes, thinner phloem, and/or less monoterpenes) seem to be the larger impediment. However, given MPB's extensive impacts, further investigation into this paper's focal contrast --- pine species versus pine volume --- is warranted.  In particular, more detailed data could be utilized to decompose the ``effect of pine volume'' into the effects of the pine size distribution, pine density, and forest connectivity. Such an analysis would help to clarify the cross-continental spread risk of this significant forest pest.

\section{Acknowledgements} 

The authors would like to thank Micah Brush, Xiaoqi Xie, and K\'evan Rastello for feedback and helpful conversation. Funding for this research has been provided through grants to the TRIA-FoR Project to ML from Genome Canada (Project No. 18202) and the Government of Alberta through Genome Alberta (Grant No.~L20TF), with contributions from the University of Alberta and fRI Research (Project No. U22004).

\section{Credits}

EJ conceived of the study, performed the analysis, and wrote the first draft; AM provided data for Figure \ref{fig:phloem_dbh} and Table \ref{tab:bolt}; CC provided pine ancestry data (i.e. the Q value raster); All authors contributed critically to the drafts and gave final approval for publication.

\section{Data availability statement}

Code and data is available for peer review (\href{https://zenodo.org/records/14553933?preview=1&token=eyJhbGciOiJIUzUxMiIsImlhdCI6MTczNTEwNjgwNSwiZXhwIjoxNzUxMzI3OTk5fQ.eyJpZCI6ImMyOGM3NGVmLWQ1ZGYtNGI1Yy04ZTFmLTg3MmJlNzQxODVhMiIsImRhdGEiOnt9LCJyYW5kb20iOiI4OTA4OGVhMDI3ODFmNzY0NTM0MTk5MzEyYjc4NWQxNCJ9.huijP08nkyPgAkq5HI1xDBgP2iOjCEMGOHCfk5fzhZuRauv6YX3R9vo-hKFsGDPvzKLs1VmyHY9i9wxX0Y5IgQ}{Zenodo files here}) and will be made publicly available if the paper is accepted.

\newpage

\begin{appendices}
\counterwithin{figure}{section}
\counterwithin{table}{section}



\section{Model justification and validation}

\subsection{Spatial thinning} \label{autocorr}

Statistical models typically assume that data points are conditionally independent, meaning residuals are uncorrelated across space and time. However, this assumption often fails when working with spatial data due to autocorrelation, where nearby observations are more similar than distant ones.

While spatial autocorrelation can sometimes be explained through predictors (such as dispersal processes or spatially correlated variables like pine volume), residual autocorrelation may persist. This conditional non-independence rarely biases parameter estimates but can make them appear falsely precise, resulting in artificially narrow confidence intervals or posterior distributions.

\begin{figure}[H]
\centering
\includegraphics[scale = 1]{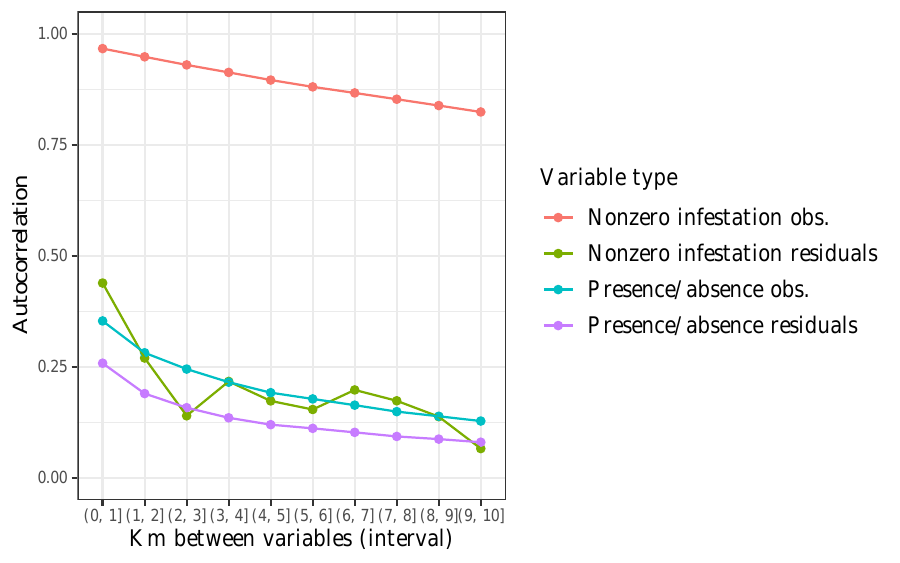}
\caption{Spatial autocorrelations of observations and residuals from model \#1 without data thinning. ``Nonzero infestations residuals'' represent the difference between the logarithm of observed and predicted infestations, for cells where some infestations are present. ``Presence/absence residuals'' is the difference between the empirical indicator variable (1 for present, 0 for absent) and the predicted probability of presence. Model \#1 explains a large fraction of the spatial autocorrelation in the number of non-zero infestations.}
\label{fig:resid_AC}
\end{figure}

In our analysis, we investigated this issue by examining spatial autocorrelation in both raw data and model residuals (Fig. \ref{fig:resid_AC}). Although our model accounted for much of the spatial structure in the number of infested trees, some conditional non-independence remained. To address this, we spatially thinned the data by selecting observations at three-kilometer intervals (corresponding to an "elbow" in the autocorrelation curve) in both the x and y directions, reducing the dataset to one-ninth of its original size. While some autocorrelation remained at 3 km, our fitted model produced marginal posterior distributions comparable to those from an unthinned model with a latent Gaussian process (Fig. \ref{fig:params_by_structure}) suggesting our thinning procedure did not lead to overconfident estimates.

\subsection{Gaussian process model} \label{GP}

Gaussian process regression, also known as kriging in geography, provides a powerful framework for handling residual spatial autocorrelation. A Gaussian process (GP) is a collection of random variables where any finite subset follows a multivariate normal distribution, completely specified by its mean and covariance functions. Gaussian processes typically model error distributions in multiple regression. In our zero-inflated negative binomial (ZINB) models, the error distribution is not clearly separable, i.e, the variance is jointly determined by the mean and size parameters. Therefore, we utilize a GP as a latent variable affecting the mean parameters of the negative binomial distribution.

The implementation of the Gaussian process modifies our original equations in the main text. For model \#1, equation \eqref{eq:pi} in the main text becomes:
\begin{equation} \label{eq:pi2}
\pi_t(x) = \begin{cases} 
\text{logit}^{-1}\left(\gamma_{0,L} + \gamma_{1,L} \log\left(B_t(x)\right) + \gamma_{\text{GP}} f_t(x)  \right) & \text{if } x \in L, \\ 
\text{logit}^{-1}\left(\gamma_{0,J} + \gamma_{1,J} \log\left(B_t(x)\right) +  \gamma_{\text{GP}} f_t(x) \right) & \text{if } x \in J,
\end{cases}
\end{equation}
Here, $\gamma_{\text{GP}}$ serves as a scale parameter, and $f_t$ represents a year-specific Gaussian process random variable drawn from a multivariate normal distribution with zero mean and covariance function:
\begin{equation}
\Sigma_t\left(x,y\right) = \exp \left(-\frac{\text{dist}\left(x,y\right)}{2 L^2} \right),
\end{equation}

The Gaussian process similarly affects the count sub-model through a different scaling factor. For model \#1, equation \eqref{eq:count} in the main text becomes:
\begin{equation} \label{eq:count2}
f(I_t(x)) = \begin{cases} 
\text{NB}\left(I_t(x) \; | \; \mu = \exp\left[\beta_{0,L} + \beta_{1,L} \log\left(B_t(x)\right)  +  \beta_{\text{GP}} f_t(x)\right], \phi = k_L \right) & \text{if } x \in L, \\
\text{NB}\left(I_t(x) \; | \; \mu = \exp\left[\beta_{0,J} + \beta_{1,J} \log\left(B_t(x)\right) +  \beta_{\text{GP}} f_t(x)\right], \phi = k_J \right) & \text{if } x \in J.
\end{cases}
\end{equation}

Gaussian processes are computationally expensive: fitting a model involves inverting the covariance matrix, generally a $O(n^3)$ operation, where $n$ is the number of pixels under consideration. However, for evenly spaced data --- like our projected pixels on a  2D lattice --- Fourier transforms can be used to invert the covariance matrix with $O(n \log n)$ operations. Using the \textit{gptools} implementation of this trick \citep{hoffmann2023scalable}, we were able to fit the model with \textit{Stan}. However, fitting the model was computationally infeasible with 1x1 km pixels, and slow with 5x5 km pixels. 

Our ultimate decision to use the non-GP model with spatial data thinning was based on several practical considerations. First, the GP and non-GP model with thinning produced remarkably similar posterior parameter distributions (Fig. \ref{fig:params_by_structure}), and this cannot be attributed to the GP model converging to the non-GP model with $L = 0$; instead, the posterior mean of the length-scale parameter was 10 km.  Second, the slow fitting of GP models hindered efficient workflow. Finally, the lack of standard methods for calculating marginal likelihoods of zero-inflated negative binomial distributions with GP latent variables complicated our model selection process, specifically when determining whether to group hybrid pines with lodgepole or jack pine in model 1 (see Section \ref{mod1}).

\subsection{Robustness analysis with model \#1} \label{robust1}

Previous research and our own experiences suggest that modeling choices can significantly impact scientific inferences \citep{draper1995assessment}. A recent statistical analysis of jack pine's suitability, conducted by Xiaoqi Xie et al. (\textit{manuscript in preparation}), found that MPB's ``reproduction rate'' (analogous to our effective brood size) was significantly smaller in jack pine. This result prompted us to examine how our modeling decisions affected our conclusions.

\begin{figure}[H]
\centering
\includegraphics[scale = 1]{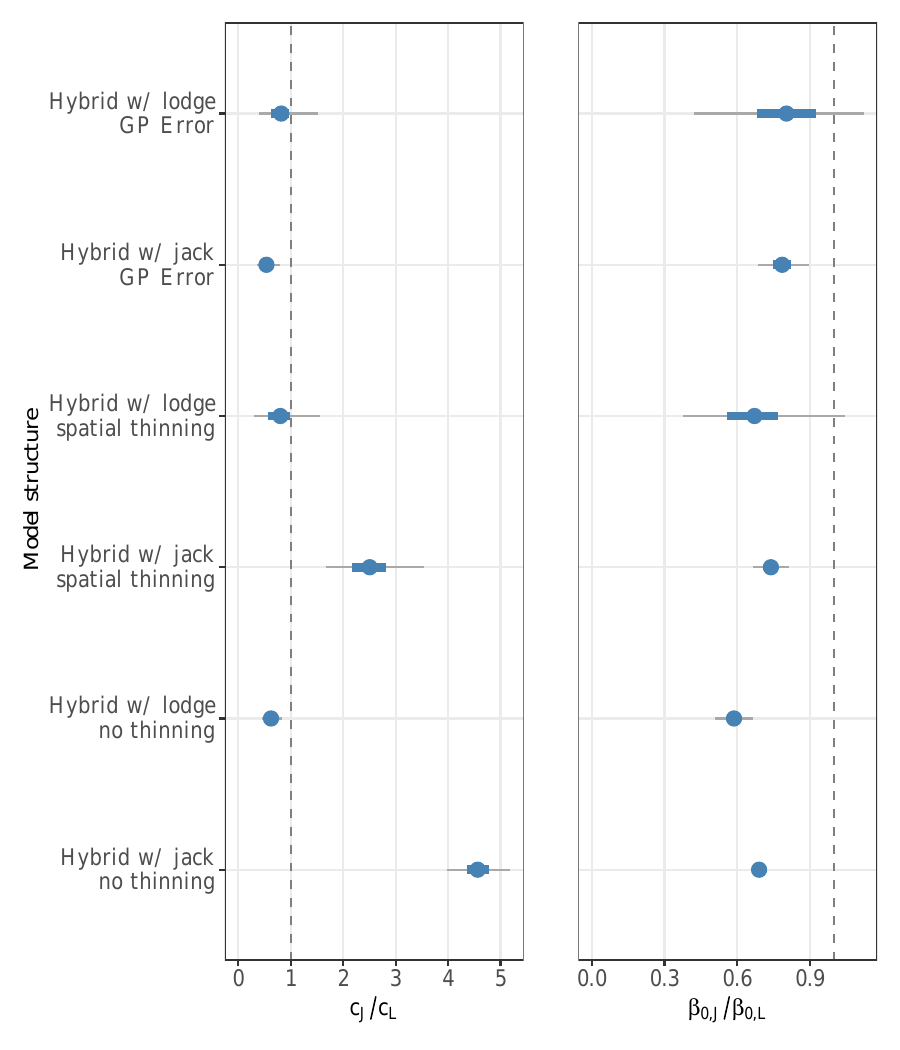}
\caption{There is substantial model uncertainty with respect to the effective brood size, but not the effective attack rate. The points are posterior means, the thick lines are 50\% credible intervals, and the thin lines are 95\% credible intervals. $c_J/c_L$ below 1 implies that MPB has a smaller effective brood size in the jack pine. $\beta_{0,J}/\beta_{0,J}$ below 1 implies that MPB has a smaller effective attack rate in jack pine. ``Hybrid with jack'' and ``Hybrid with lodge'' refers to the grouping of hybrid pixels (i.e., $0.1 \leq Q < 0.9$) with pure jack pine or pure lodgepole pine pixels, for the purpose of fitting model \#1, as described in Section \ref{mod1} of the main text. ``Spatial thinning'' refers to the removal of all data except for the intersection of every third row and column, as described in Section \ref{data_prep} of the main text. ``GP'' refers to the Gaussian process models described in Appendix \ref{GP}.}
\label{fig:params_by_structure}
\end{figure}

We fit six different variants of model \#1 in the main text, and found that inferences about the effective brood size face substantial within and between-model uncertainty, which may alternatively be described respectively as parameter uncertainty and structural uncertainty (Fig. \ref{fig:params_by_structure}). On the other hand, all models consistently showed that the effective attack rate is lower in jack pine, as evidenced by  $\beta_{0,J}/\beta_{0,J} < 1$. Though other parameters affect the effective attack rate (e.g. $\gamma_{0,J}$, see Section \ref{mod1} for more info) exploratory simulations revealed that the intercept parameters capture the overall pattern of attack rates across pine species.

When applying the same data pre-processing methods as Xie et al. --- grouping hybrid and lodgepole pixels without spatial data thinning --- we obtained similar results showing significantly lower effective brood size in jack pine. However, this posterior distribution is falsely narrow due to conditional non-independence of the data (Fig. \ref{fig:params_by_structure}). Adjusting for residual autocorrelation through spatial data thinning or a Gaussian process latent variable revealed that the effective brood size quotient $c_J/c_L$ has a posterior distribution that substantially overlaps with values above and below 1, with a posterior mean close to unity (see rows 1 and 3 in Fig. \ref{fig:params_by_structure}).

When hybrid and jack pine pixels are grouped together, we observe more extreme model uncertainty, with posterior distributions of $c_J/c_L$ that are either completely above or below unity. We compared the hybrid+jack vs. hybrid+lodgepole groupings with approximate leave-one-out cross-validation, as implemented by the \textit{loo} \textit{R Package} \citep{vehtari2019loo}. The expected log predictive density (ELPD) served as the predictive metric. We found that models grouping hybrid and lodgepole pixels performed significantly better according to cross validation; specifically, $\mathrm{E}\left[\Delta ELPD\right] \gg 2 \cdot SE(\Delta ELPD)$. Full details can be found in the file {\fontfamily{qcr}\selectfont scripts/model\_diagnostics.Rmd} in the supplementary materials. We limited our comparisons to models with spatial data thinning, as models without data thinning would produce artificially narrow $SE(\Delta ELPD)$, and ELPD computation for Gaussian process models is computationally impractical; one would need to use a computationally expensive method to calculate the marginal likelihood (e.g., stepping stone sampling, bridge sampling) for every MCMC sample.

Our robustness analysis reveals substantial structural uncertainty in the effective brood size, but not the effective attack rate. While data pre-processing choices significantly impact our conclusions, we were able to identify a small subset of good models. Models grouping hybrid and lodgepole pixels show superior predictive performance. Among these, the two models that also account for spatial autocorrelation produced similar posterior distributions, where $c_J/c_L$ was uncertain but close to 1 on average.

\subsection{Model \#2 with interaction effects} \label{mod2_int}

We extended Model \#2 to include all possible two-way interactions between the predictors. The model structure remains similar but with additional interaction terms in both the presence and count components. The probability of observing a non-zero number of infestations becomes
\begin{equation}
\text{logit}\left(\pi\right) =
\gamma_{0} + \gamma_{B} B^* + \gamma_{Q} Q^* + \gamma_{V} V + \gamma_{BV} B^*V + \gamma_{BQ} B^*Q^* + \gamma_{QV} Q^* V,
\end{equation}
where $B^*$, $Q^*$, and $V$ represent the standardized beetle pressure (log-transformed), pine ancestry, and pine volume (log-transformed) predictors, respectively. 

The mean of the negative binomial count distribution similarly becomes
\begin{equation}
\mu = \exp\left[\beta_{0} + \beta_{B} B^* + \beta_{Q} Q^* + \beta_{V} V + \beta_{BV} B^*V + \beta_{BQ} B^*Q^* + \beta_{QV} Q^* V \right]
\end{equation}

Table \ref{tab:pars3} reveals several notable interaction effects. In the presence/absence component, there is a substantial negative interaction between pine ancestry and pine volume ($\gamma_{QV} = -0.42$), suggesting that the effect of pine ancestry on infestation probability decreases in areas with higher pine volume. In the count component, we found smaller positive interactions between beetle pressure and both pine volume ($\beta_{BV} = 0.10$) and pine ancestry ($\beta_{BQ} = 0.15$). 

\begin{figure}[H]
\centering
\includegraphics[scale = 1]{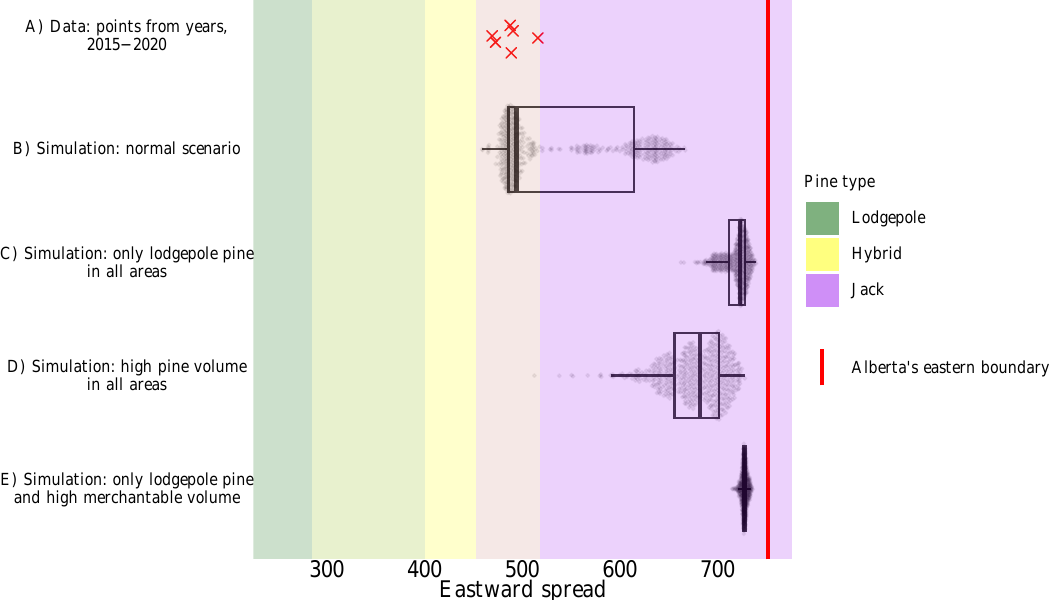}
\caption{A variant of model \#2 with interaction effects supports a main result of this paper: mountain pine beetle's slow spread is primarily due to some species-level property of jack pine, and secondarily due to low pine volumes in eastern Alberta. The x-axis shows eastward spread as measured by the 99th percentile of infestation distance, from 2015--2020, along the projection line (see Fig \ref{fig:study_areas}).}
\label{fig:sim_vol_boxplot_int}
\end{figure}

We used counterfactual simulations to show how these interaction effects work together. Figure \ref{fig:sim_vol_boxplot_int} shows these results, which parallel the non-interaction analysis presented in Figure \ref{fig:sim_vol_boxplot} of the main text. These simulations support a main conclusion of this paper: both pine volume and pine ancestry matter, but pine ancestry matters more. 

\begin{table}[ht]
\centering
\begin{tabular}{llllll}
  \hline
Parameter & Short description & Mean & SD & $\text{CI}_{2.5\%}$ & $\text{CI}_{97.5\%}$ \\ 
  \hline
$\gamma_{0}$ & Intercept, presence & -3.8 & 0.083 & -4.0 & -3.7 \\ 
  $\gamma_{B}$ & Beetle pressure, presence & 1.6 & 0.078 & 1.4 & 1.7 \\ 
  $\gamma_{Q}$ & Pine ancestry, presence & 0.32 & 0.090 & 0.14 & 0.50 \\ 
  $\gamma_{V}$ & Pine volume, presence & 1.2 & 0.11 & 0.98 & 1.4 \\ 
  $\gamma_{BV}$ & Beetle pressure \& pine volume interaction, presence & 0.031 & 0.071 & -0.11 & 0.17 \\ 
  $\gamma_{BQ}$ & Beetle pressure \& pine ancestry interaction, presence & -0.10 & 0.083 & -0.27 & 0.057 \\ 
  $\gamma_{QV}$ & Pine ancestry \& pine volume interaction, presence & -0.42 & 0.12 & -0.66 & -0.18 \\ 
  $\beta_{0}$ & Intercept, count & 1.8 & 0.099 & 1.6 & 2.0 \\ 
  $\beta_{B}$ & Beetle pressure, count & 0.037 & 0.062 & -0.087 & 0.16 \\ 
  $\beta_{Q}$ & Pine ancestry, count & 0.54 & 0.12 & 0.29 & 0.76 \\ 
  $\beta_{V}$ & Pine volume, count & 0.010 & 0.16 & -0.31 & 0.33 \\ 
  $\beta_{BV}$ & Beetle pressure \& pine volume interaction, count & 0.10 & 0.040 & 0.024 & 0.18 \\ 
  $\beta_{BQ}$ & Beetle pressure \& pine ancestry interaction, count & 0.15 & 0.068 & 0.024 & 0.29 \\ 
  $\beta_{QV}$ & Pine ancestry \& pine volume interaction, count & -0.0045 & 0.19 & -0.37 & 0.37 \\ 
  $k$ & Dispersion parameter, count & 0.42 & 0.022 & 0.38 & 0.47 \\
   \hline
\end{tabular}
\caption{Parameter estimates model \#2 with interaction effects. Recall that the predictors $B$ $V$, and $Q$ have been standardized so that the coefficients can be interpreted as predictor importance.} 
\label{tab:pars3}
\end{table}

\subsection{Model-fitting details} \label{model_details}

We examined a standard suite of diagnostics \citep[Ch. 6]{gelman2014bayesian}. to ensure that the Markov Chains had converged to a unique posterior distribution and were sampling efficiency. All models under consideration passed these diagnostics successfully. Specifically, we confirmed that $R < 1.1$ for all parameters (indicating proper chain mixing), the effective sample size per iteration exceeded 0.001 (demonstrating efficient sampling), the energy Bayesian fraction of missing information (E-BFMI) was below 0.2 (suggesting appropriate model specification), and the proportion of divergent trajectories remained well below 1\% (indicating unbiased estimation). The complete diagnostic analysis can be found in the supplementary files, specifically {\fontfamily{qcr}\selectfont scripts/model\_diagnostics.Rmd \& \selectfont scripts/stan\_utility.R}

To evaluate the influence of our prior distributions on the posterior estimates, we calculated the posterior contraction:
\begin{equation}
\text{post. contraction} = 1 - \frac{ \mathbb{V}{\mathrm{post}}} { \mathbb{V}{\mathrm{prior}}}.
\end{equation}
This metric quantifies how informative the data are relative to the prior for each parameter. All parameters showed posterior contraction greater than 0.99, with the exception of parameter $c_J$, which exhibited posterior contraction of 0.83.  This is not unexpected, given that the contraction measures the relative informativeness of the data, and the fact that there are few infestations in jack pine. The posterior contraction analysis can also be found in the supplementary files: {\fontfamily{qcr}\selectfont  scripts/model\_diagnostics.Rmd}.

\subsection{Additional tables and figures}

\begin{figure}[H]
\centering
\includegraphics[scale = 1]{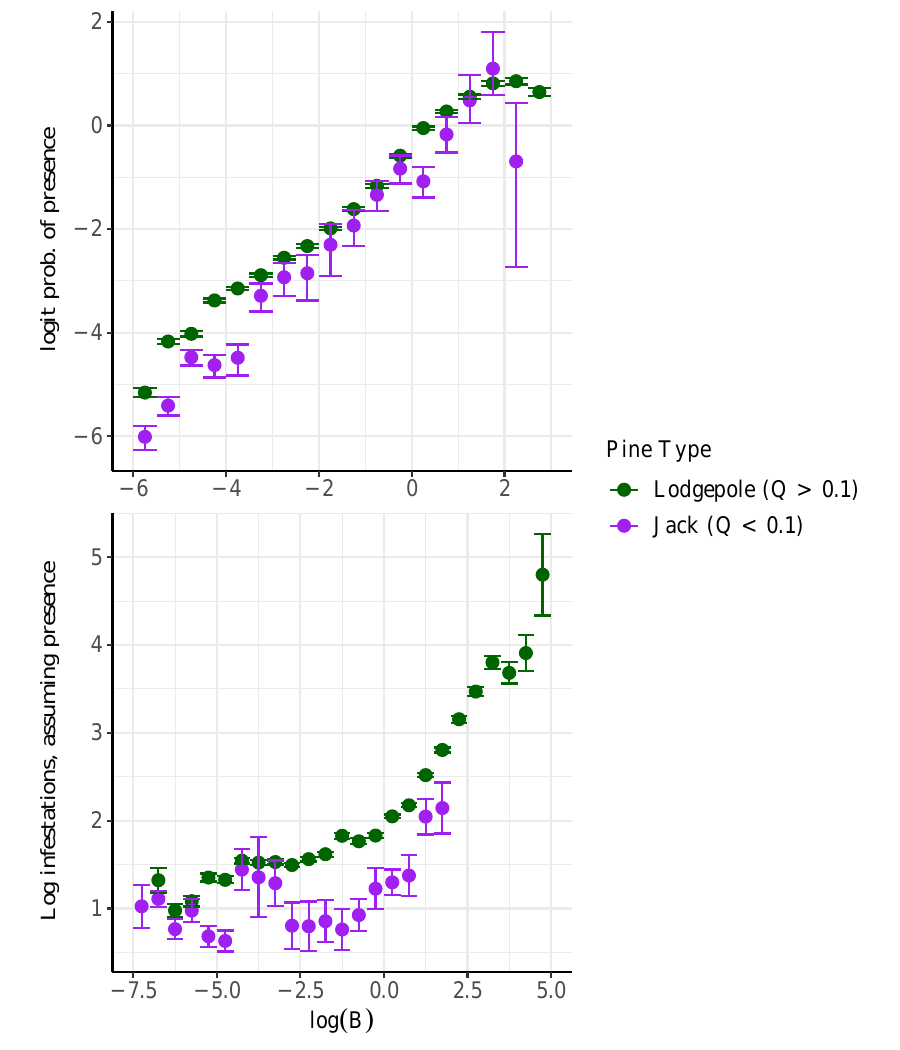}
\caption{Graphical justification of model \#1. The logit-scale probability of infestations within a 1$\times$1 km pixel, and the logarithm of total infestations within a 1$\times$1 km pixel (assuming that some infestations are present), are approximately linear with respect to the logarithm of beetle pressure. Points and errorbars show the mean $\pm$ 1 standard error across pixels within evenly spaced intervals of log beetle pressure, $\log(B)$.}
\label{fig:mod1_just}
\end{figure}

\begin{figure}[H]
\centering
\includegraphics[scale = 1]{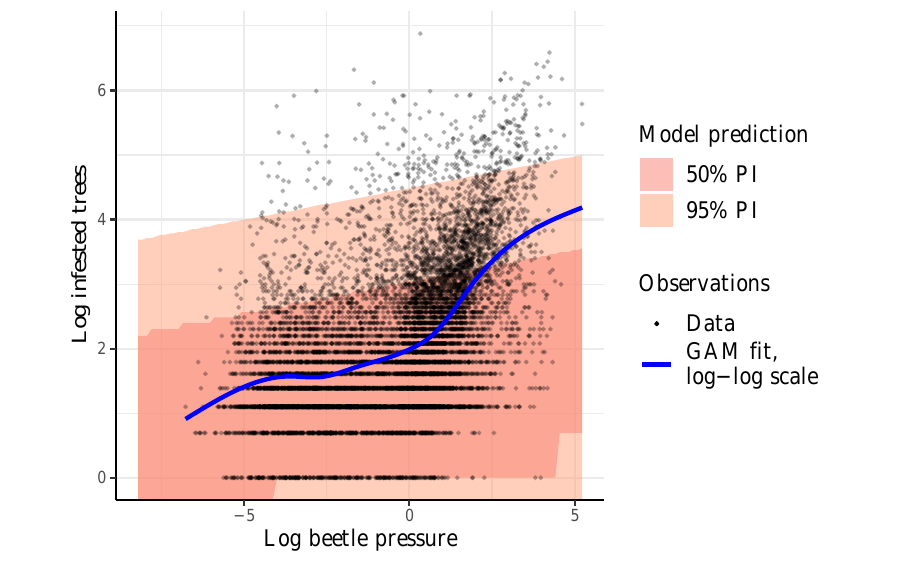}
\caption{Observations and predictive intervals (PIs) for the negative binomial count sub-model of model \#1. Each point represents the logarithm of the total number of infested trees within a 1$\times$1 km pixel for a particular year.  Pixel-year combinations with zero infestations are not shown.}
\label{fig:NB_pred_vs_obs}
\end{figure}

\begin{figure}[H]
\centering
\includegraphics[scale = 1]{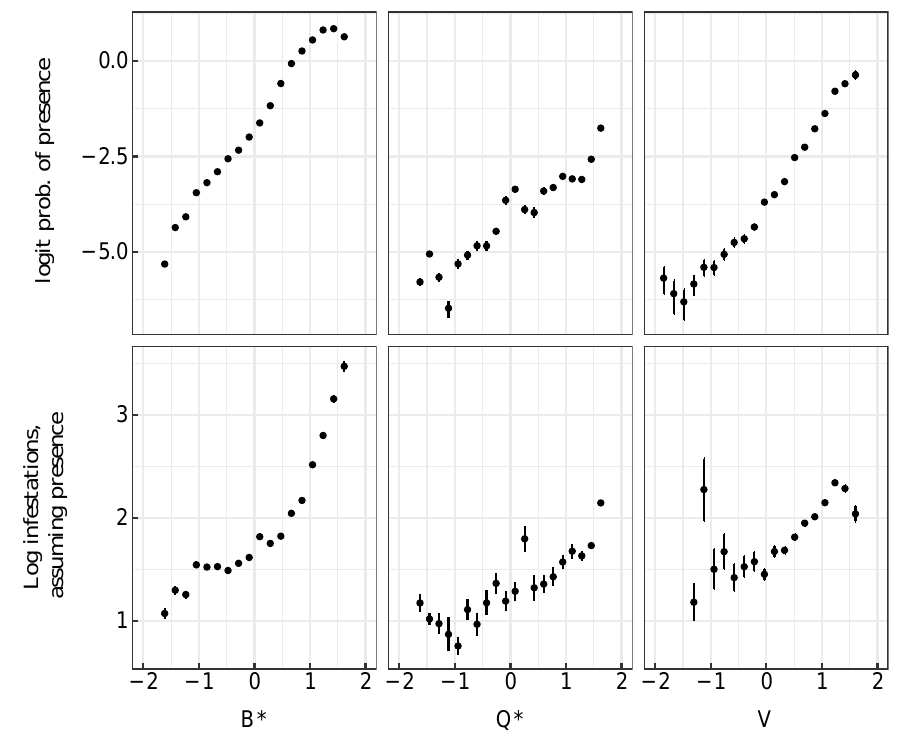}
\caption{Graphical justification of model \#2. The logit-scale probability of infestations, and the logarithm of infestations (assuming 1 or more infestations) are approximately linear with respect to all three predictors. Points and errorbars show the mean $\pm$ 1 standard error across pixels within evenly spaced intervals of the predictors .}
\label{fig:mod2_just}
\end{figure}

\begin{figure}[H]
\centering
\includegraphics[scale = 1]{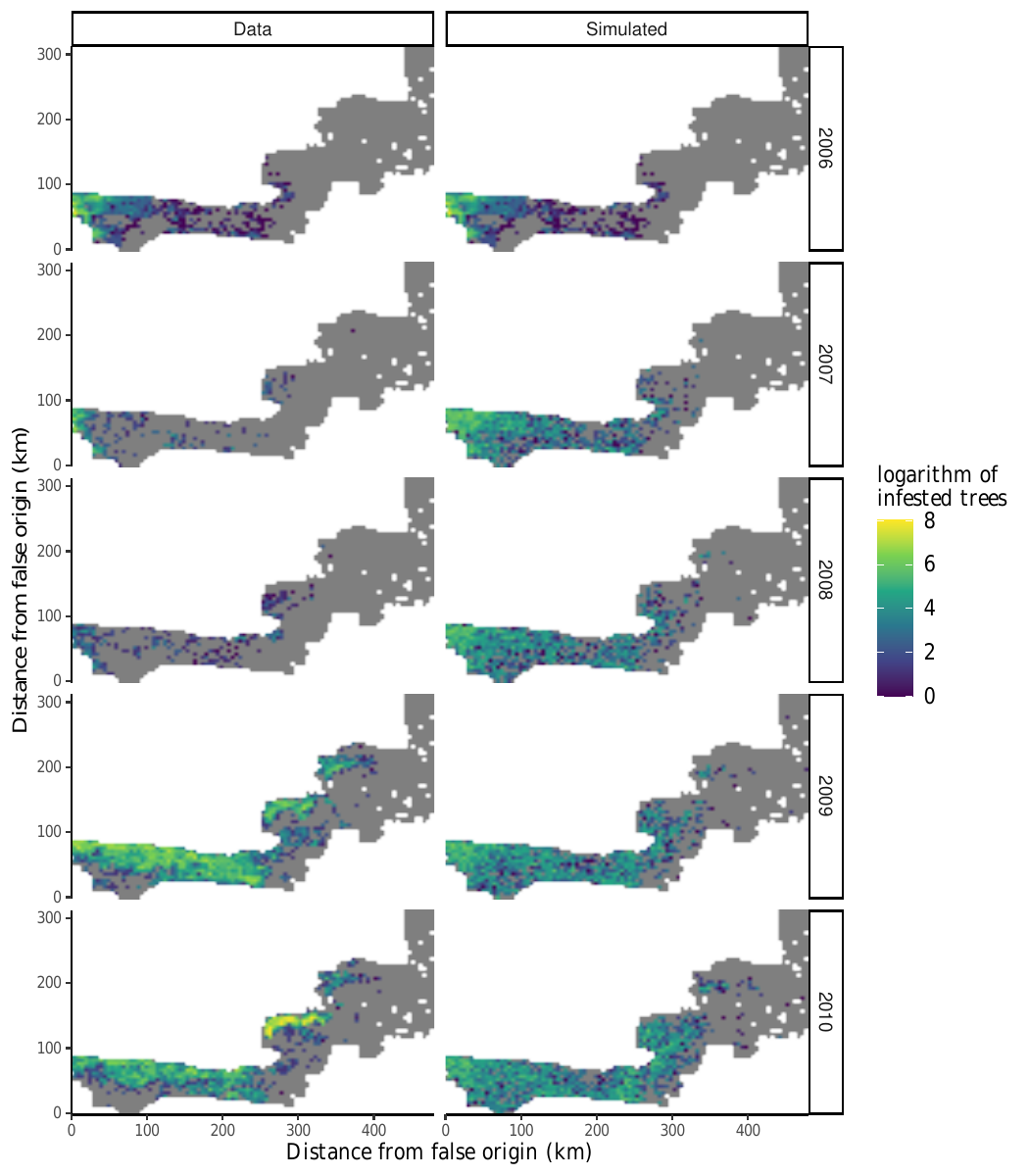}
\caption{Model limitations: simulations of model \#1 starting in 2006 do not capture the high number of infestations in 2009 \& 2010, nor the patchy distribution of infestations. Data (real and simulated) are shown here in 5x5 km pixels.}
\label{fig:sim_map}
\end{figure}

\begin{table}[ht]
\centering
\begin{tabular}{llllll}
  \hline
Parameter & Short description & Mean & SD & $\text{CI}_{2.5\%}$ & $\text{CI}_{97.5\%}$ \\ 
  \hline
$c_L$ & Effective brood size, lodgepole/hybrid & 1 & 0 & N/A & N/A \\ 
  $c_J$ & Effective brood size, jack & 0.80 & 0.33 & 0.30 & 1.6 \\ 
  $\gamma_{0,L}$ & Presence logit intercept, lodgepole/hybrid & -0.091 & 0.053 & -0.19 & 0.014 \\ 
  $\gamma_{0,J}$ & Presence logit intercept, jack & -1.7 & 0.47 & -2.7 & -0.79 \\ 
  $\gamma_{1,L}$ & Presence logit slope, lodgepole/hybrid & 0.64 & 0.014 & 0.62 & 0.67 \\ 
  $\gamma_{1,J}$ & Presence logit slope, jack & 0.62 & 0.086 & 0.46 & 0.79 \\ 
  $\beta_{0,L}$ & Abundance log intercept, lodgepole/hybrid & 2.7 & 0.037 & 2.6 & 2.8 \\ 
  $\beta_{0,J}$ & Abundance log intercept, jack & 1.8 & 0.46 & 1.0 & 2.8 \\ 
  $\beta_{1,L}$ & Abundance log slope, lodgepole/hybrid & 0.10 & 0.0087 & 0.086 & 0.12 \\ 
  $\beta_{1,J}$ & Abundance log slope, jack & 0.074 & 0.063 & 0.0023 & 0.23 \\ 
  $k_{L}$ & Dispersion param, lodgepole/hybrid & 0.40 & 0.020 & 0.37 & 0.45 \\ 
  $k_{J}$ & Dispersion param, jack & 0.67 & 0.33 & 0.24 & 1.5 \\ 
   \hline
\end{tabular}
\caption{Parameter estimates model \#1. Note that the effective brood size in lodgepole pine is fixed at $c_L=1$ in order for other parameters to be identifiable. } 
\label{tab:pars1}
\end{table}

\begin{table}[ht]
\centering
\begin{tabular}{llllll}
  \hline
Parameter & Short description & Mean & SD & $\text{CI}_{2.5\%}$ & $\text{CI}_{97.5\%}$ \\ 
  \hline
$\gamma_{0}$ & Intercept, presence & -3.8 & 0.069 & -3.9 & -3.7 \\ 
  $\gamma_{B}$ & Beetle pressure, presence & 1.5 & 0.039 & 1.4 & 1.6 \\ 
  $\gamma_{Q}$ & Pine ancestry, presence & 0.16 & 0.070 & 0.028 & 0.30 \\ 
  $\gamma_{V}$ & Pine volume, presence & 0.93 & 0.055 & 0.83 & 1.0 \\ 
  $\beta_{0}$ & Intercept, count & 1.7 & 0.082 & 1.5 & 1.8 \\ 
  $\beta_{B}$ & Beetle pressure, count & 0.25 & 0.024 & 0.20 & 0.29 \\ 
  $\beta_{Q}$ & Pine ancestry, count & 0.58 & 0.087 & 0.41 & 0.75 \\ 
  $\beta_{V}$ & Pine volume, count & 0.12 & 0.057 & 0.0027 & 0.23 \\ 
  $k$ & Dispersion parameter, count & 0.42 & 0.021 & 0.38 & 0.46 \\ 
   \hline
\end{tabular}
\caption{Parameter estimates model \#2.  Recall that the predictors $B^*$ $V$, and $Q^*$ have been standardized so that the coefficients can be interpreted as predictor importance.} 
\label{tab:pars2}
\end{table}

\begin{figure}[H]
\centering
\includegraphics[scale = 1]{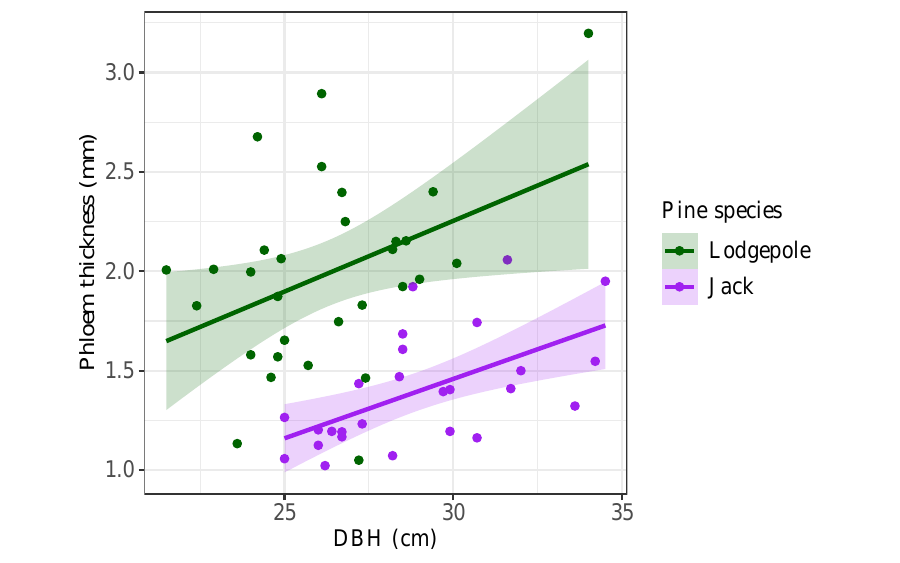}
\caption{Jack pine has thinner phloem than lodgepole pine at the same diameter at breast height (DBH). Data from \citet{musso2023pine}, chapters 2 \& 3.}
\label{fig:phloem_dbh}
\end{figure}

\begin{figure}[H]
\centering
\includegraphics[scale = 1]{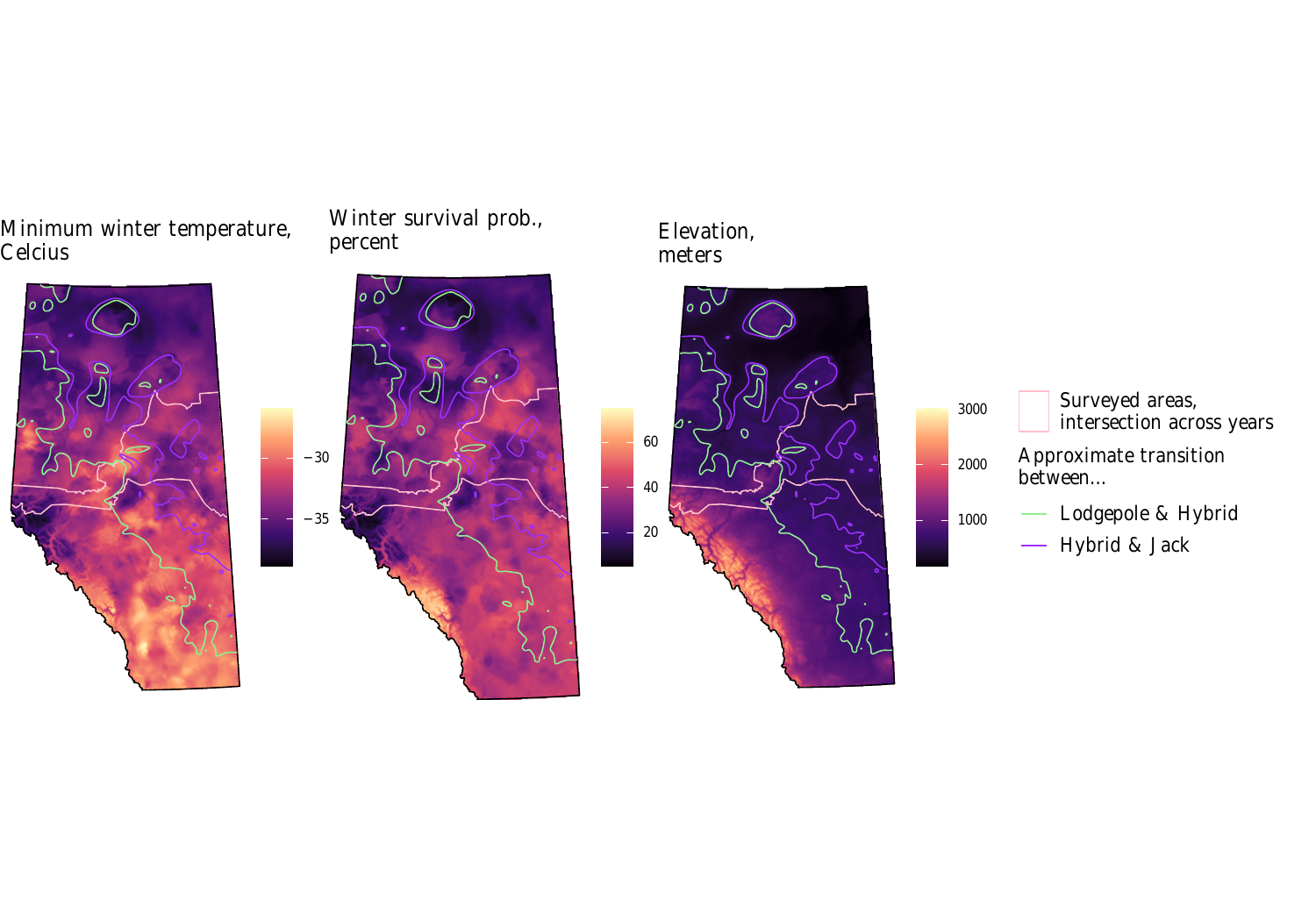}
\caption{The thermal regime in jack pine forests is suitable for MPB development and survival, partly because eastern Alberta has a lower elevation.}
\label{fig:covariate_maps_jack_temp}
\end{figure}

\begin{figure}[H]
\centering
\includegraphics[scale = 1]{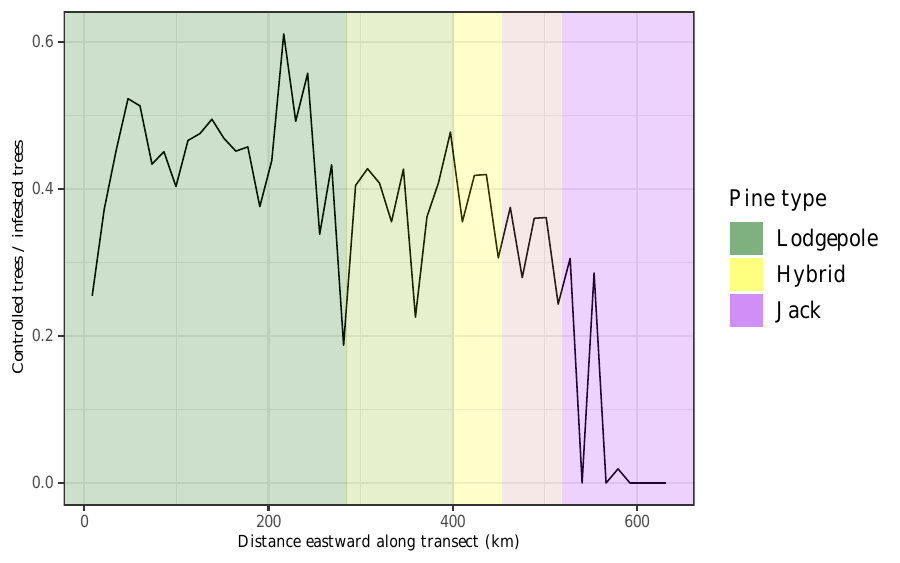}
\caption{A higher proportion of infested trees were controlled in western Alberta, compared to eastern Alberta. This figure was created by projecting infestations and controlled trees within the consistently surveyed area polygon, onto to the projection line, as in Figure \ref{fig:data_spread_w_inset} in the main text.}
\label{fig:cntrl_1D}
\end{figure}


\begin{figure}[H]
\centering
\includegraphics[scale = 1]{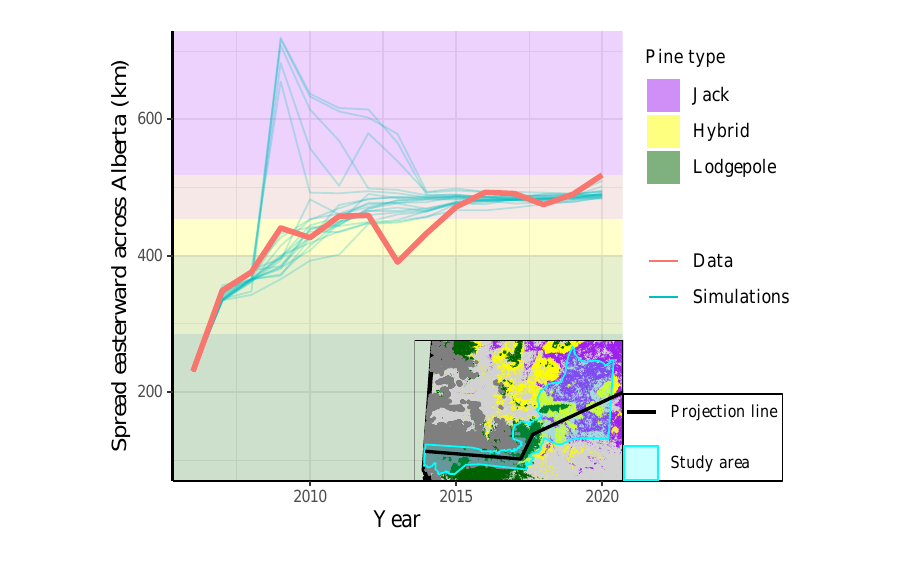}
\caption{Model \#1 accurately predicts the observed pattern MPB's decelerating spread. The peak of some simulations in 2009 comes from the fact that the proportion of controlled trees was low in 2009, putatively due to long-distance dispersal from British Columbia into western Alberta \citep{carroll2017assessing}.}
\label{fig:sim_spread_w_inset}
\end{figure}

\section{The relationship between beetle pressure and brood density} \label{disk}

To examine the relationship between beetle pressure and brood size, we analyzed Alberta's MPB \textit{disk data} \citep{government2016mountain}. Collected by Alberta Agriculture and Forestry, this dataset spans from 2008 to 2016, covers over 1000 sites across Alberta, and contains nearly 10,000 trees. The sampling method involves extracting four bark-covered sapwood disks from attacked trees (up to 20 trees per site). Entrance holes and various life stages are counted. Like \citet{goodsman2018effect}, we excluded trees that contained no larvae or pupae (whether living or dead) from our analysis, since the absence of any life stages indicates these trees were not successfully attacked. The brood size is measured simply as the total number of living MPB (all stages), and brood density per unit of the tree's surface area is easily calculated given that each disk has a 10.6 cm diameter.

As a proxy for beetle pressure, we counted the previous year's infestations within a local neighborhood of each disk sample. For example, if disks are processed during the spring of 2009, then beetle pressure is the number of red-topped trees in in the autumn of 2008. To demonstrate that our results are robust to subjective modeling decisions, we utilized circular neighborhoods of two different sizes: one with a 100 m radius, and another with a 500 m radius.

\begin{figure}[H]
\centering
\includegraphics[scale = 0.8]{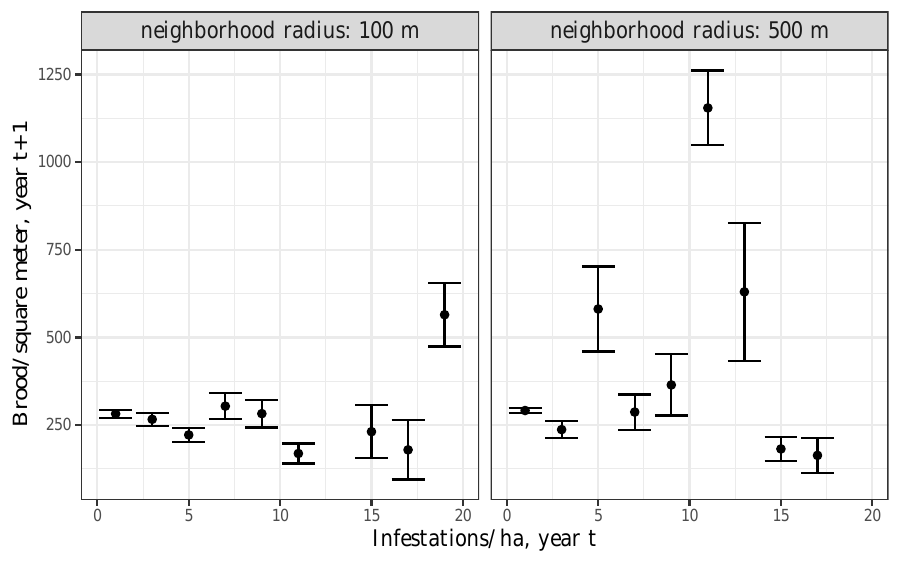}
\caption{The relationship between beetle pressure and brood density, specifically beetles per square meter of tree surface area.}
\label{fig:brood_vs_bp}
\end{figure}

Figure \ref{fig:brood_vs_bp} shows the lack of a clear relationship between beetle pressure and brood density. We categorized observations into bins based on the number of infestations per hectare, using a bin width of 2. Additional plots using different bin widths (not shown here) revealed similar patterns.

\end{appendices}

\bibliographystyle{apalike}
\bibliography{mpb_refs.bib}

\end{document}